\documentclass[useAMS,usenatbib]{mn2e}

\usepackage{times,graphicx,amssymb,natbib}

\newcommand{\degrees}{^{\circ}}
\newcommand{\msol}{M_{\rm \odot}}
\newcommand{\mjup}{M_{\rm Jup}}
\newcommand{\rjup}{R_{\rm Jup}}
\newcommand{\mearth}{M_{\rm \oplus}}

\title[Exomoon Climate Models with Planetary Irradiation]{The Effect of Planetary Illumination on Climate Modelling of Earthlike Exomoons}
\author[Duncan Forgan and Vergil Yotov]{Duncan Forgan $^{1,2}$\thanks{E-mail:dhf@roe.ac.uk} and Vergil Yotov$^{1}$ \\
$^{1}$Scottish Universities Physics Alliance (SUPA), Institute for Astronomy, University of Edinburgh, Blackford Hill, Edinburgh, EH9 3HJ, Scotland, UK \\
$^{2}$UK Centre for Astrobiology, School of Physics and Astronomy, University of Edinburgh \\
}

\begin{document}

\date{Accepted}

\pagerange{\pageref{firstpage}--\pageref{lastpage}} \pubyear{}

\maketitle

\label{firstpage}

\begin{abstract}

\noindent  From analytical studies of tidal heating, eclipses and planetary illumination, it is clear that the exomoon habitable zone (EHZ) - the set of moon and host planet orbits that permit liquid water on an Earthlike moon's surface - is a manifold of higher dimension than the planetary HZ.  

This paper outlines the first attempt to produce climate models of exomoons which possess all the above sources and sinks of energy.  We expand on our previous 1D latitudinal energy balance models (LEBMs), which follow the evolution of the temperature on an Earthlike moon orbiting a Jupiterlike planet, by adding planetary illumination.

We investigate the EHZ in four dimensions, running two separate suites of simulations.  The first investigates the EHZ by varying the planet's orbit, keeping the moon's orbit fixed, to compare the EHZ with planetary habitable zones.  In general, planetary illumination pushes EHZs slightly further away from the star.  

Secondly, we fix the planet's orbit and vary the moon's orbit, to investigate the circumplanetary inner habitable edge.  We demonstrate that an outer edge can exist due to eclipses (rather than merely orbital stability), but this edge may be pushed outwards when the effect of the carbonate-silicate cycle is taken into account.

\end{abstract}


\section{Introduction}

% Introduce exomoons
% Introduce previous attempts to model exomoon climates
% Introduce importance of factors such as magnetic shielding, planetary irradiation, potentially causing a greenhouse
% Introduce climate modelling techniques and previous uses

\noindent There are currently no confirmed detections of extrasolar moons (exomoons).  The only moons known to humanity thus far are those which reside in our own Solar System.  However, there are several proposed methods of detecting exomoons within the ability of current instrumentation, amongst which are: exoplanet transit timing and duration variations \citep{Simon2007, Kipping2009, Heller2014b};  microlensing \citep{Liebig2010}; and direct imaging, provided that the moon is strongly heated by tidal forces \citep{Peters2013}.  The Hunt for Exomoons with Kepler team \citep{Kipping2012} are now attempting to make such a detection using transit data from the Kepler Space Telescope. 

Any first detection is likely to lie at the very edge of instrumental ability, and as such the primary objective of current missions is to establish upper limits on the occurrence rate of satellites in their samples (cf \citealt{Weidner2010,Montalto2012,Kipping2013a, Kipping2014}).  As observers continue to reduce these upper limits, it is reasonable to expect that detections of exomoons will soon follow.  If these exomoons are Earthlike, and the host planet orbits in the habitable zone (HZ) of their parent star, then it is possible that the moons themselves are habitable (given other factors which we will discuss below).  

The HZ concept was originally created to investigate planetary habitability \citep{Huang1959}.  Assuming that a planet has Earthlike properties such as mass, atmospheric composition and surface water (amongst others), the HZ is calculated by modelling how stellar radiative flux interacts with the planet's atmosphere, and what equilibrium temperature that planet subsequently adopts.  If the temperature is conducive to the surface water being liquid, that planet is said to be within the habitable zone.  As it is determined in the first instance by radiative flux, the planetary HZ is a sensitive function of distance from the star, and is a spherical annulus.  Planets beyond the inner edge of the HZ will generally suffer water loss by photolysis and hydrogen escape after a runaway greenhouse effect; planets beyond the outer edge of the HZ experience runaway glaciation as $CO_2$ clouds form (see e.g. \citealt{Kasting_et_al_93} and \citealt{Kopparapu2013} for more detail).  

Moons possess sources and sinks of energy that planets do not. If their orbit around the planet is eccentric, tidal forces can dissipate in the moon's interior, releasing heat.  This mechanism allows icy Solar System moons such as Europa \citep{Melosh2004} and Enceladus \citep{plume_enceladus, Iess2014} to possess liquid oceans beneath ice crusts; equally, it can produce extensive volcanism, on moons such as Io \citep{Peale1979,Veeder2012}.  The magnitude of such tidal heating, dependent on the gravitational field of the planet and the moon's interaction with it, has important consequences for habitability.  

The planet's radiation field also plays an important role.  This includes both the thermal radiation the planet emits and the starlight reflected by the planet, thanks to its non-zero albedo.  If the planet is tidally locked to the star, then the moon will experience a variation in the planet's thermal flux as it orbits it, periodically forcing the climate in ways that planets are unlikely to experience.  The spectrum of the planet's radiation is also important - strong EUV or X-Ray radiation can result in catastrophic atmosphere loss \citep{Kaltenegger2000}.  These processes are in turn linked to the planet's magnetic field, another factor that can either promote or destroy exomoon habitability depending on the magnetosphere's ability to shield the moon from high energy photons \citep{Heller2013a}.

Finally, moons are likely to experience frequent eclipses of starlight due to the host planet, which act as an effective sink of energy.  The extent to which eclipses screen stellar flux from the moon is a sensitive function of the moon's orbit about the planet \citep{Heller2012}.

In short, it is clear that the planetary HZ and the equivalent lunar HZ will differ in their spatial extent, as well as in the factors that determine that spatial extent.  \citet{Reynolds1987} demonstrated that Europa presented a viable niche for terrestrial marine life, and proposed a circumplanetary habitable zone determined by tidal heating.  \citet{Williams1997b} considered potential sites for the first exomoons, albeit from the then-limited sample of known exoplanets, where they considered the risks of the moon becoming tidally locked, the potentially hazardous local radiation environment and the (over-)abundance of volatiles.

\citet{Scharf2006} used a more populous exoplanet sample to consider potential habitable exomoon hosts.  Estimating that around 30\% of the population could host icy moons amenable to subsurface oceans via tidal heating, the main consequence was an extension of the combined exoplanet/exomoon HZ by a factor of almost 2 (if one is willing to accept heating levels significantly larger than Io, which is likely to have other negative consequences). 

\citet{Heller2013} constructed models of the insolation received by an exomoon in orbit of a tidally locked exoplanet.  These models calculated the flux as a function of moon surface longitude and latitude, taking into account direct stellar radiation, eclipses, planetary radiation and tidal heating, demonstrating that there is indeed an inner circumplanetary ``habitable edge'', beyond which moons are heated too strongly (either by tidal heating or planetary illumination) to avoid becoming uninhabitable.  

Later modelling \citep{Hinkel2013} has shown that, much as planets need not spend their entire orbit within the HZ to be habitable themselves \citep{Williams_and_Pollard_02,Kane2012,Kane2012c}, exoplanets hosting exomoons need not spend their entire orbit within the planetary HZ for the exomoon to be habitable, depending on the moon's heat redistribution efficiency.

In a previous paper \citep{Forgan_moon1} we developed a latitudinal energy balance model (LEBM) that described the one-dimensional temperature evolution of an Earthlike moon in orbit of a Jupiterlike planet, which in turn orbits a Sunlike star.  Rather than finding an equilibrium surface temperature through analytical calculation, we allow the temperature to evolve due to the radiative flux of the star, tidal heating, eclipses, the moon's albedo, the transfer of heat through the atmosphere, and its loss via infrared radiation.  

In this paper, we improve our LEBM, by including the planet's thermal blackbody radiation and starlight reflected by the planet.  In section \ref{sec:LEBM} we describe the model and the improvements we have made.  Section \ref{sec:results} describes the results of two separate suites of simulations, where we investigate the exomoon habitable zone (EHZ) in terms of 

\begin{enumerate}
\item the planet semimajor axis and eccentricity, and
\item the moon semimajor axis and eccentricity
\end{enumerate}

In section \ref{sec:discussion} we discuss the limitations of the model and the implications of the results, and in section \ref{sec:conclusions} we summarise the work.

\section{Latitudinal Energy Balance Modelling} \label{sec:LEBM}

% Diagram of the simulation setup
% Description of the 1DEBM

\subsection{Simulation Setup}

\noindent We proceed in a manner similar to that of \citet{Forgan_moon1}.  The star mass is fixed at $M_*=1 \msol$, the mass of the host planet at $M_p = 1 \mjup$, and the mass of the moon at 1$\mearth$.  This system has been demonstrated to be dynamically stable on timescales comparable to the Solar System lifetime \citep{Barnes2002}.  

We specify the planet's orbit around the star using standard orbital elements: the semi-major axis $a_p$ and eccentricity $e_p$.  The moon's orbit around the planet is given by its semi-major axis $a_m$, eccentricity $e_m$ and the inclination relative to the planet's orbital plane is fixed at  $i_m=0^\circ$.  The orbital longitudes of the planet and moon are defined such that $\phi_{p}=\phi_{m}=0$ corresponds to the x-axis.  We assume $a_m$ describes the average distance of the moon from the planet, rather than the average distance of the moon to the barycentre of the planet-moon system.  This approximation is satisfactory given the relatively low mass of the moon relative to the planet.

Much of the model setup is as described in our previous paper on exomoon LEBMs \citep{Forgan_moon1}.  We repeat this description below, and show where planetary irradiation is added to the model.

\subsection{A One Dimensional Latitudinal Energy Balance Model with Tidal Heating and Planetary Irradiation}

\noindent The core equation of standard exoplanet LEBMs is a diffusion equation, which we modify for our purposes here:

\begin{equation} 
C \frac{\partial T}{\partial t} - \frac{\partial }{\partial x}\left(D(1-x^2)\frac{\partial T}{\partial x}\right) = (S+S_p)\left[1-A(T)\right] - I(T), 
\end{equation}

\noindent where $T=T(x,t)$ is the temperature at time $t$, $x = \sin \lambda$, and $\lambda$ is the latitude (between $-90\degrees$ and $90\degrees$).  This equation is evolved with the boundary condition $\frac{dT}{dx}=0$ at the poles.  The $(1-x^2)$ term is a geometric factor, arising from solving the diffusion equation in spherical geometry.

$C$ is the effective heat capacity of the atmosphere, $D$ is a diffusion coefficient that determines the efficiency of heat redistribution across latitudes, $S$ is the insolation received from the star, $S_p$ is the insolation received from the planet, $I$ is the atmospheric infrared cooling and $A$ is the moon's albedo.  In the above equation, $C$, $S$, $S_p$, $I$ and $A$ are functions of $x$ (either explicitly, as $S$ is, or implicitly through $T(x,t)$).   

$D$ is a free parameter which we must tune to reproduce the climate of an Earth-like exoplanet at 1 AU around a star of $1 \msol$, with diurnal period of 1 day.   Judicious selection of $D$ allows us to reproduce the seasonally averaged temperature profile measured on Earth (see e.g. \citealt{North1981, Spiegel_et_al_08}).  Planets that rotate rapidly experience inhibited latitudinal heat transport, due to Coriolis forces truncating the effects of Hadley circulation (cf \citealt{Farrell1990, Williams1997a}).  We follow \citet{Spiegel_et_al_08} by scaling $D$ according to:

\begin{equation} 
D=5.394 \times 10^2 \left(\frac{\omega_d}{\omega_{d,\oplus}}\right)^{-2},\label{eq:D}
\end{equation}

\noindent where $\omega_d$ is the rotational angular velocity of the planet, and $\omega_{d,\oplus}$ is the rotational angular velocity of the Earth.   This is a simplified expression, which fixes the atmospheric pressure of the planet as well as the various partial pressures of gases such as water vapour and $CO_2$.  It is an approximate but limited description of the effects of rotation - for example it does not account for Hadley circulation, which could be achieved by allowing $D$ to vary with latitude (e.g. \citealt{Vladilo2013}), but allows for rapid computation without severely compromising the model's accuracy.

The diffusion equation is solved using a simple explicit forward time, centre space finite difference algorithm.  A global timestep was adopted, with constraint

\begin{equation}
\delta t < \frac{\left(\Delta x\right)^2C}{2D(1-x^2)}.  
\end{equation}

\noindent This timestep constraint ensures that the first term on the left hand side is always larger than the second term, preventing the diffusion term from setting up unphysical temperature gradients.  The parameters are diurnally averaged, i.e. a key assumption of the model is that the moons rotate sufficiently quickly relative to their orbital period around the insolation source.  As the primary insolation source is the star, and the moon rotates relative to the star on the timescale of days as it orbits the planet, this approximation is broadly satisfied, even in cases where we might expect pseudo-synchronous rotation.

The atmospheric heat capacity depends on what fraction of the moon's surface is ocean, $f_{ocean}$, what fraction is land $f_{land}=1.0-f_{ocean}$, and what fraction of the ocean is frozen $f_{ice}$:

\begin{equation} 
C = f_{land}C_{land} + f_{ocean}\left[(1-f_{ice})C_{ocean} + f_{ice} C_{ice}\right]. 
\end{equation}

\noindent The heat capacities of land, ocean and ice covered areas are 

\begin{equation} 
C_{land} = 5.25 \times 10^9 $ erg cm$^{-2}$ K$^{-1},
\end{equation}

\begin{equation} C_{ocean} = 40.0C_{land},\end{equation}
\begin{equation} C_{ice} = \left\{
\begin{array}{l l }
9.2C_{land} & \quad \mbox{263 K $< T <$ 273 K} \\
2C_{land} & \quad \mbox{$T<263$ K} \\
0.0 & \quad \mbox{$T>$ 273 K}.\\
\end{array} \right. 
\end{equation}

\noindent These parameters assume a wind-mixed ocean layer of 50m \citep{Williams1997a}.  Increasing the assumed depth of this layer would increase $C_{ocean}$ (see e.g. \citealt{North1983} for details). We use the following infrared cooling function:

\begin{equation} 
I(T) = \frac{\sigma_{SB}T^4}{1 +0.75 \tau_{IR}(T)}, 
\end{equation}

\noindent where the optical depth of the atmosphere 

\begin{equation} 
\tau_{IR}(T) = 0.79\left(\frac{T}{273\,\mathrm{K}}\right)^3. 
\end{equation}

\noindent The albedo function is

\begin{equation} 
A(T) = 0.525 - 0.245 \tanh \left[\frac{T-268\, \mathrm K}{5\, \mathrm K} \right]. 
\end{equation}

\noindent This produces a rapid shift from low albedo ($\sim 0.3$) to high albedo ($\sim 0.75$) as the temperature drops below the freezing point of water, producing highly reflective ice sheets.  Figure 1 of \citet{Spiegel_et_al_08} demonstrates how this shift in albedo affects the potential for global energy balance, and that for planets in circular orbits, two stable climate solutions arise, one ice-free, and one ice-covered. Note that we do not consider clouds in this model, which could modify both the albedo and optical depth of the system significantly.  Also, we assume that both stellar and planetary flux are governed by the same albedo, which in truth is not likely to be the case (see Discussion).

The stellar insolation flux $S$ is a function of both season and latitude.  At any instant, the bolometric flux received at a given latitude at an orbital distance $r$ is

\begin{equation}
S = q_0\cos Z \left(\frac{1 AU}{r}\right)^2,
\end{equation}

\noindent where $q_0$ is the bolometric flux received from the star at a distance of 1 AU, and $Z$ is the zenith angle:

\begin{equation} 
q_0 = 1.36\times 10^6\left(\frac{M_*}{\msol}\right)^4 \mathrm{erg \,s^{-1}\, cm^{-2}} 
\end{equation}

\begin{equation} 
\cos Z = \mu = \sin \lambda \sin \delta + \cos \lambda \cos \delta \cos h. 
\end{equation} 

\noindent We have assumed a simple main sequence scaling for the luminosity (see e.g. \citealt{Prialnik}). $\delta$ is the solar declination, and $h$ is the solar hour angle.  The moon's obliquity $\delta_0$ is fixed at 23.5 degrees relative to its orbit around the planet. The solar declination is calculated as:

\begin{equation} 
\sin \delta = -\sin \delta_0 \cos(\phi_{*m}-\phi_{peri,m}-\phi_a), 
\end{equation}

\noindent where $\phi_{*m}$ is the current orbital longitude of the moon \emph{relative to the star}, $\phi_{peri,m}$ is the longitude of periastron, and $\phi_a$ is the longitude of winter solstice, relative to the longitude of periastron.   We set $\phi_{peri,m}=\phi_a=0$ for simplicity. 

We must diurnally average the solar flux:

\begin{equation} 
S = q_0 \bar{\mu}. 
\end{equation}

\noindent This means we must first integrate $\mu$ over the sunlit part of the day, i.e. $h=[-H, +H]$, where $H$ is the radian half-day length at a given latitude.  Multiplying by the factor $H/\pi$ (as $H=\pi$ if a latitude is illuminated for a full rotation) gives the total diurnal insolation as

\begin{equation} 
S = q_0 \left(\frac{H}{\pi}\right) \bar{\mu} = \frac{q_0}{\pi} \left(H \sin \lambda \sin \delta + \cos \lambda \cos \delta \sin H\right). \label{eq:insol}
\end{equation}

\noindent The radian half day length is calculated as

\begin{equation} 
\cos H = -\tan \lambda \tan \delta. 
\end{equation}

\subsubsection{Eclipses}

\noindent To simulate the effect of eclipses of the star by the planet, we set $S$ to zero at all latitudes whenever an eclipse is detected.  The detection algorithm relies on the angle $\alpha$ between the vector connecting the moon and planet, $\mathbf{s}$, and the vector connecting the moon and the star $\mathbf{s}_{*}$:

\begin{equation}
\cos \alpha = \mathbf{\hat{s}.\hat{s}_*} 
\end{equation}

\noindent It is straightforward to show that an eclipse is in progress if

\begin{equation}
\left|s_*\right| \sin \alpha < R_p
\end{equation}

\noindent We do not model the eclipse ingress and egress, which would allow $S$ to decrease and increase monotonically, and instead simply set $S$ to zero at any point during an eclipse.  As our simulations address the case where all orbits are coplanar, we are in effect considering the case where eclipse probabilities are maximised, and hence the effective sink of eclipses is at its maximum strength.

\subsubsection{Tidal Heating}

\noindent In the interest of computational expediency, we make a simple approximation for tidal heating, by firstly assuming the tidal heating per unit area is \citep{Peale1980,Scharf2006}:

\begin{equation} 
\left(\frac{dE}{dt}\right)_{tidal} = \frac{21}{38}\frac{\rho^2_m R^{5}_m e^2_m}{\Gamma Q}\left(\frac{GM_p}{a^3_m}\right)^{5/2} 
\end{equation}

\noindent where $\Gamma$ is the moon's elastic rigidity (which we assume to be uniform throughout the body), $R_m$ is the moon's radius, $\rho_m$ is the moon's density, $M_p$ is the planet mass, $a_m$ and $e_m$ are the moon's orbital semi-major axis and eccentricity (relative to the planet), and $Q$ is the moon's tidal dissipation parameter.  We assume terrestrial values for these parameters, hence $Q=100$, $\Gamma=10^{11} \,\mathrm{dyne \, cm^{-2}}$ (appropriate for silicate rock) and $\rho_m=5 \, \mathrm{g \, cm^{-3}}$.

We assume that this heating occurs uniformly across the moon's surface.  This is very much an approximation - indeed, the multi-dimensional nature of tidal heating prohibits latitudinal models from improving much on approximations such as this. Exomoon habitability studies typically assume tidal heating is uniformly distributed throughout the body (see e.g. \citealt{Heller2013}).

\subsubsection{Planetary Illumination}

\noindent We implement planetary illumination in the same manner as \citet{Heller2013}.  We assume that the planet surface has dayside temperature $T_{day}$ and nightside temperature 

\begin{equation} 
T_{night} = T_{day}-dT,
\end{equation}

\noindent where we fix the temperature difference $dT=100$ K.  The total flux received by the moon from the planet is a combination of the planet's thermal blackbody flux and the starlight it reflects:

\begin{equation}
S_p(t)=f_t(t)+f_r(t).
\end{equation}

\noindent The thermal flux $f_t$ as a function of the moon's surface longitude $\varphi$ and latitude $\lambda$ is:

\begin{equation}
f_t(t)= \frac{R_p^2 \sigma_{SB}}{a_{m}^2} \cos \varphi \cos \lambda \left( T_{day}^4 \Xi(t)+T_{night}^4 (1-\Xi(t)) \right),
\end{equation}

\noindent And the reflected flux $f_r$ is:

\begin{equation}
f_r(t)=\frac{L_*}{4\pi r_{p*}^2} \frac{R_p^2 \pi \alpha_p}{a_{m}^2} 
	\cos \varphi \cos \lambda \Xi(t).
\end{equation}

\noindent Where $\alpha_p$ is the albedo of the planet, $r_{p*}$ is the distance between the planet and the star, and $\Xi$ describes what fraction of the dayside of the planet is visible to the moon.  We calculate $\Xi$ in the same manner as shown in Appendix B of \citet{Heller2013} (they use the symbol $\xi$, which we have reserved for habitability calculations, see following section).  The fraction of visible dayside depends on the location of the moon's projection on the planetary surface (the subplanetary point), and the planet's own substellar point.  The substellar point lies along the planet's orbital plane, with an azimuth depending on the planet's true anomaly $\nu_{*p}$, and the subplanetary point is given by spherical co-ordinates $(\vartheta,\Phi)$.  Hence

\begin{equation}
\Xi(t)=\frac{1}{2} \left\lbrace 1+\cos \left(\vartheta(t)\right) \cos \left(\Phi(t)-\nu_{*p}(t)\right) \right\rbrace,
\end{equation}

where 
\begin{equation}
\Phi(t)=2*\arctan \left( \frac{s_y(t)}{s_x(t)+\sqrt{s_x^2(t)+s_y^2(t)}} \right)	
		\end{equation}	
\begin{equation}
\vartheta(t)=\frac{\pi}{2}-\arccos \left( \frac{s_y(t)}{\sqrt{s_x^2(t)+s_y^2(t)+s_z^2(t)}} \right)	
		\end{equation}		

and $\mathbf{s} = (s_x,s_y,s_z)$ is the vector connecting the planet and moon centres.  As the LEBM is averaged over the planet's surface longitudes, we integrate over $\varphi$ and substitute:

\begin{equation}
\cos \varphi \rightarrow \int \cos\varphi  d\varphi = 2
\end{equation}

\noindent To find the dayside and nightside temperature of the planet, we stipulate thermal equilibrium, and balance the absorbed and emitted radiation impinging on the planet:

\begin{equation}
T^4_{day} + T^4_{night} - 2*T^4_* \frac{(1-\alpha_p)R_*^2}{r_{p*}^2} = 0
\end{equation}

\noindent Substituting for $T_{night}$ using the fixed value of $dT$ gives a quartic equation for $T_{day}$, which we solve iteratively.  For simplicity, we fix the albedo of the planet at 0.3.

\subsubsection{Habitability Indices \label{sec:habindex}}

\noindent We calculate habitability indices in the same manner as most groups do \citep{Spiegel_et_al_08, Vladilo2013, Forgan2014}.  The habitability function $\xi$\footnote{This function is often labelled $\eta$ - we choose $\xi$ to avoid confusion with the frequency of Earth-like/habitable planets, often denoted as $\eta$} is:

\begin{equation} 
\xi(\lambda,t) = \left\{
\begin{array}{l l }
1 & \quad \mbox{273 K $< T(\lambda,t) <$ 373 K} \\
0 & \quad \mbox{otherwise}. \\
\end{array} \right. \end{equation}

\noindent We then average this over latitude to calculate the fraction of habitable surface at any timestep:

\begin{equation} 
\xi(t) = \frac{1}{2} \int_{-\pi/2}^{\pi/2}\xi(\lambda,t)\cos \lambda \, d\lambda. 
\end{equation}

\noindent Each simulation is allowed to evolve until it reaches a steady or quasi-steady state, and the final ten years of climate data are used to produce a time-averaged value of $\xi(t)$, $\bar{\xi}$, and the sample standard deviation, $\sigma_{\xi}$.  We use these two parameters to classify each simulations as follows:

\begin{enumerate}
\item \emph{Habitable Moons} - these moons possess a time-averaged $\bar{\xi}>0.1$, and $\sigma_{\xi} < 0.1\bar{\xi}$, i.e. the fluctuation in habitable surface is less than 10\% of the mean.
\item \emph{Hot Moons} - these moons have average temperatures above 373 K across all seasons, and are therefore conventionally uninhabitable, and $\bar{\xi} <0.1$.
\item \emph{Snowball Moons} - these moons have undergone a snowball transition to a state where the entire moon is frozen, and are therefore conventionally uninhabitable.  As with hot moons, we require $\bar{\xi}<0.1$ for the moon to be classified as a snowball, but given the nature of the snowball transition as it is modelled here, these worlds typically have $\bar{\xi}=0$.
\item \emph{Transient Moons} - these moons possess a time-averaged $\bar{\xi}>0.1$, and $\sigma_{\xi} > 0.1\bar{\xi}$, i.e. the fluctuation in habitable surface is greater than 10\% of the mean.
\end{enumerate}

\section{Results}\label{sec:results}

\noindent As is now well understood, the habitable zone for exomoons (when expressed in terms of orbital elements) is a manifold of higher dimensions than that of the habitable zone for exoplanets.  In the following sections we consider how the habitable zone varies as a function of $a_p$, $e_p$, $a_m$ and $e_m$.  We consider $a_m$ in units of the Hill Radius:

\begin{equation} 
R_H  = a_p\left(\frac{M_p}{3M_*}\right)^{1/3}.
\end{equation}

We run two separate suites of simulations.  In the first, we hold $a_m$ and $e_m$ constant, and investigate the exomoon habitable zone (EHZ) in terms of $a_p$ and $e_p$.  In the second, we hold $a_p$ and $e_p$ constant, and investigate the EHZ in terms of $a_m$ and $e_m$.  In all runs, we demand that $a_m < 0.3 R_H$ for orbital stability \citep{Holman1999,Barnes2002}.  This is quite a conservative constraint - we note that \citet{Domingos2006} found limits of $a_m < 0.49 R_H$ from dynamical simulations using the restricted three-body problem in the case of circular orbits.  We do not consider the effect of planet or moon eccentricity on orbital stability, but we do note its importance.

\subsection{The Exomoon Habitable Zone in terms of Planetary Orbit}

\subsubsection{The Habitable Zone for an Earthlike Planet}

\begin{figure*}
\begin{center}$
\begin{array}{cc}
\includegraphics[scale = 0.4]{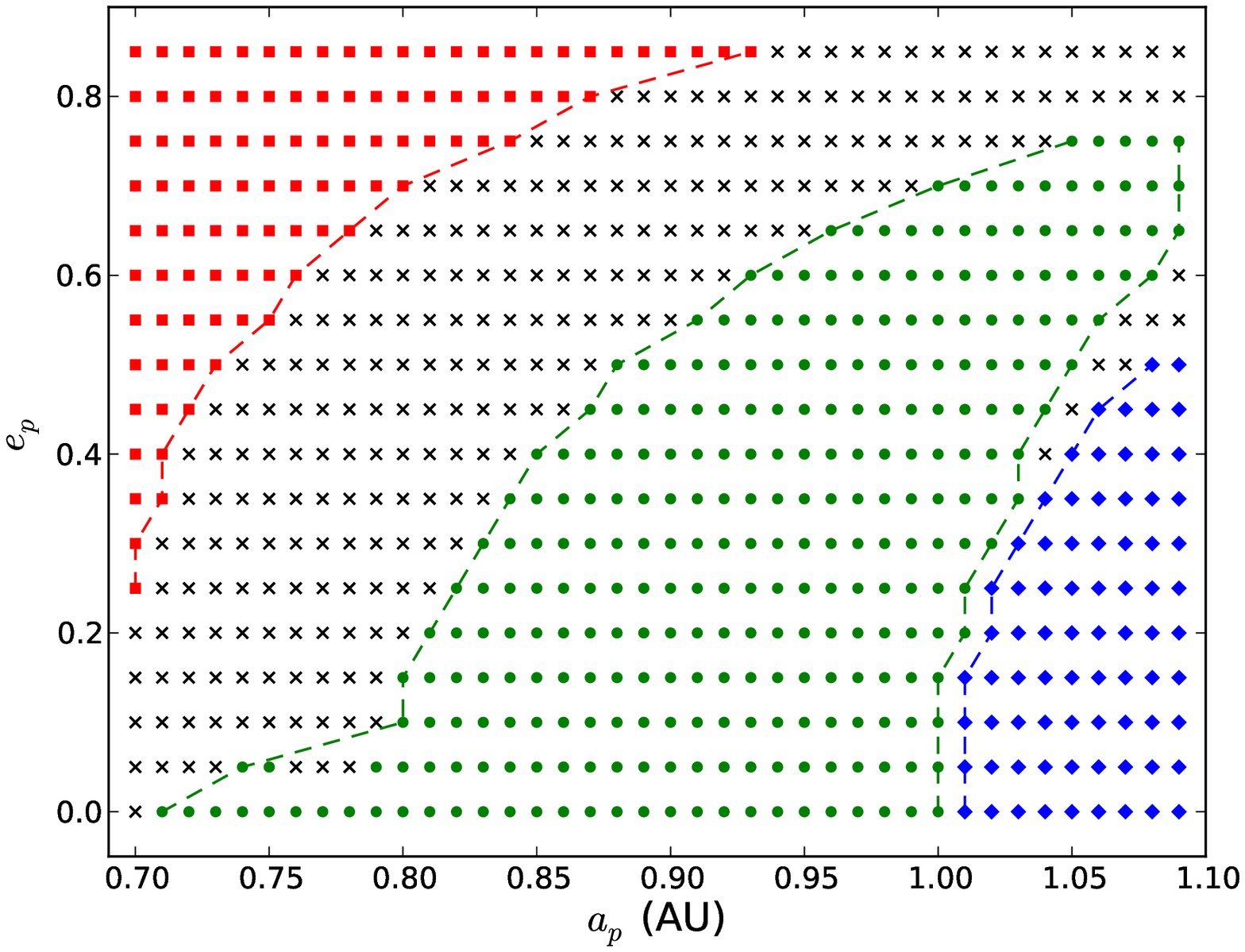} &
\includegraphics[scale = 0.4]{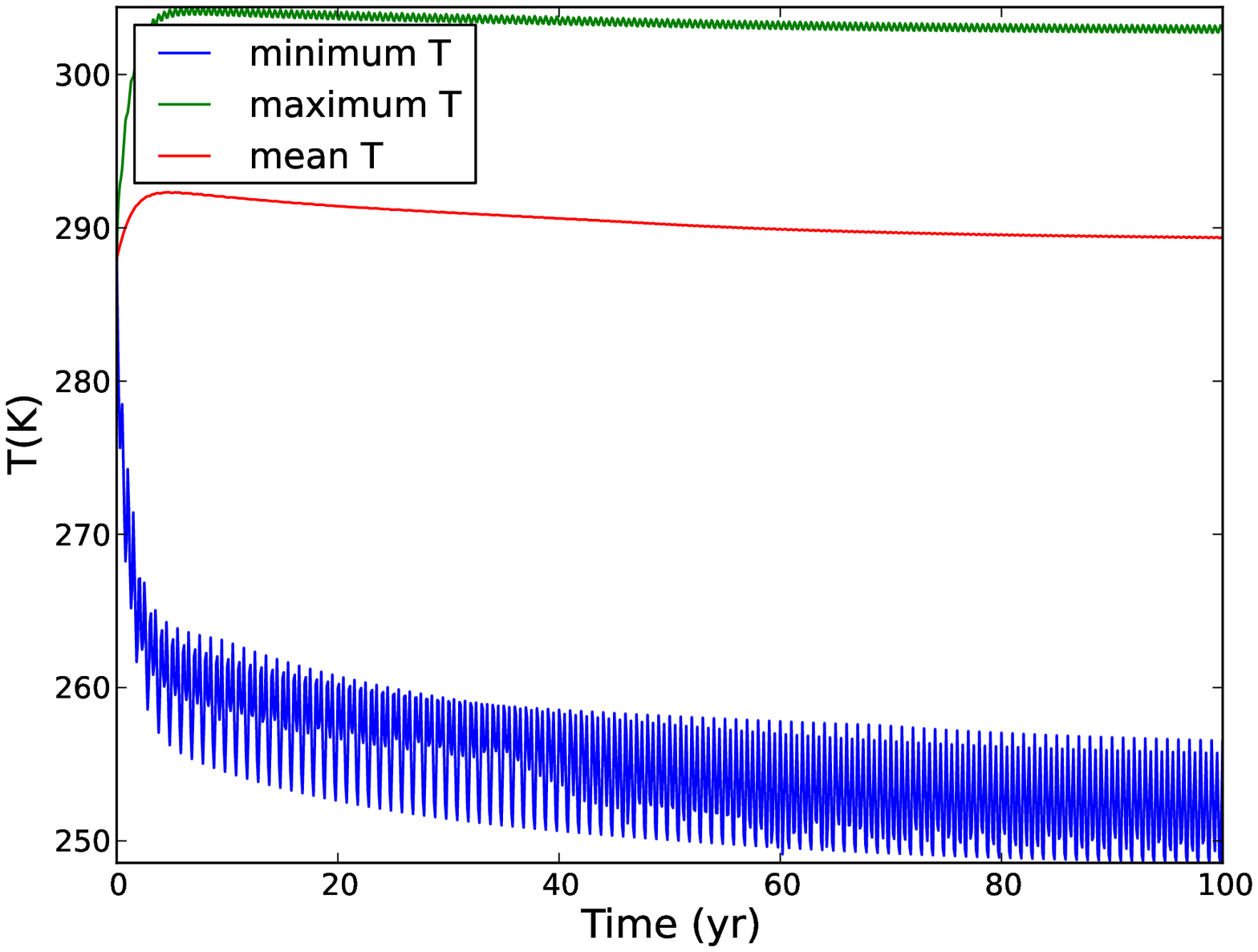} \\
\end{array}$
\caption{Left:The habitable zone for an Earth-like planet around a Sun-like star, as calculated from a LEBM using the classification system outlined in this paper. Each point represents a simulation run with these parameters, and the colour of the point indicates its outcome.  Red squares designate hot planets with little to no habitable surface; blue diamonds produce cold planets with little to no habitable surface; green circles represent warm planets with at least ten percent of the surface habitable and low seasonal fluctuations; crosses represent warm planets with high seasonal fluctuations. Right: The global minimum, maximum and mean temperatures on an Earth-like planet as a function of time at $a_p=1AU$, $e_p=0$, with initial temperature $T=288$ K throughout. \label{fig:earth_ae_space}}
\end{center}
\end{figure*}

\noindent The classification system used here distinguishes between worlds that are habitable with a small variation in surface liquid water over time, and worlds which are habitable with a large variation in surface liquid water over time.  As in \citet{Forgan_moon1}, we use the above classification system to generate a habitable zone for an Earthlike planet around a Sunlike star.  We run LEBM models at a series of planet semimajor axes $a_p$ and eccentricities $e_p$, with the results displayed in the left panel of Figure \ref{fig:earth_ae_space}.  The right panel shows the global minimum, maximum and mean temperatures on the Earth-like planet with $a_p=1AU$, $e_p=0$.

The model does not contain a carbonate-silicate cycle, unlike e.g. \citealt{Williams1997a}, which modified the atmospheric $CO_2$ pressure in accordance with temperature dependent weathering rates.  Lower temperatures produce lower weathering rates, and as a result the atmospheric $CO_2$ (produced by volcanic outgassing) cannot be sequestered.  Therefore, cooler planets can be expected to have higher atmospheric concentrations of $CO_2$, boosting the greenhouse effect and moving the outer edge of the HZ to higher semi-major axes than we see in Figure \ref{fig:earth_ae_space}  (cf \citealt{Kasting_et_al_93}).

The extension of the HZ at low $e_p$ to semi-major axes as low as 0.7 AU is a reflection of our (fairly lenient) criterion for habitability - namely, that 10\% of the planet's surface remains habitable over a ten year period, with a standard deviation less than 10\% of the mean habitable area.  As $e_p$ is increased, $\sigma_\xi$ increases quickly, producing planets which are habitable on a seasonal basis only.  In the case of planets that are snowballs at low $e_p$, e.g. at semimajor axes above 1.05 au, increasing eccentricity can increase both $\bar{\xi}$ and $\sigma_{\xi}$, with the net effect being that these worlds are classified as transiently habitable.  Equally, if considering planets at a fixed eccentricity of above 0.4, then increasing $a_p$ from 1.0 to 1.1 AU decreases $\bar{\xi}$ while keeping $\sigma_{\xi}$ approximately constant, which changes their classification from a warm planet to a transient one, before finally succumbing to a snowball transition at large enough $a_p$.  These phenomena are direct results of our classification system.

If we are to compare to traditional habitability studies, then we should infer that the inner edge of the HZ is at approximately 0.8 AU for $e_p=0$.  Equally, many of the transient classifications in this study would normally have been considered to be uninhabitable (as many of these simulations undergo seasonal periods when the habitable surface fraction is close to zero, but this is not sufficient to label them as hot or cold planets).  We should bear this in mind as we consider the habitable zone for exomoons in the following sections.

\subsubsection{Exomoon Habitability in the Absence of Tidal Heating}

\begin{figure*}
\begin{center}$
\begin{array}{cc}
\includegraphics[scale = 0.4]{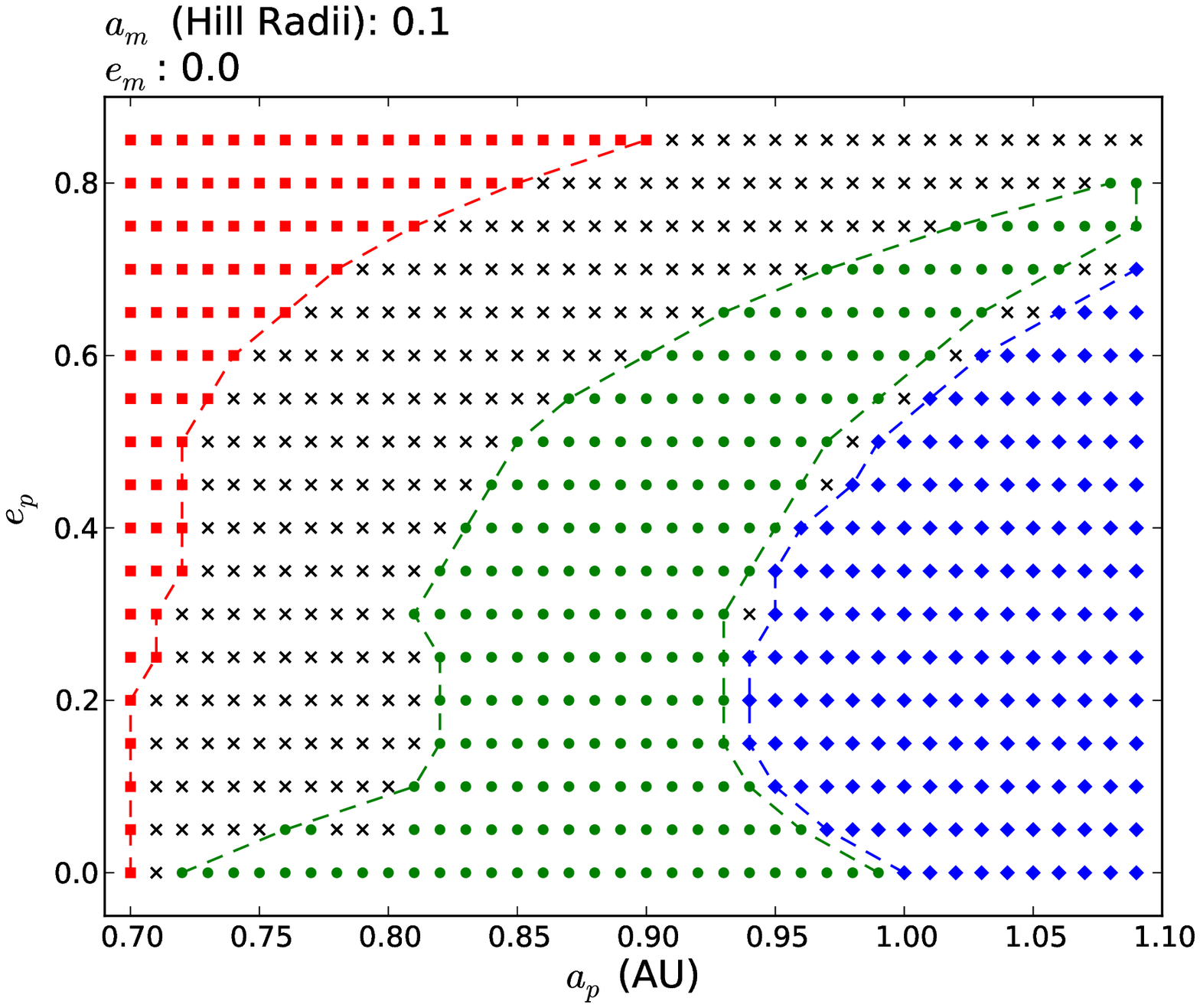} &
\includegraphics[scale = 0.4]{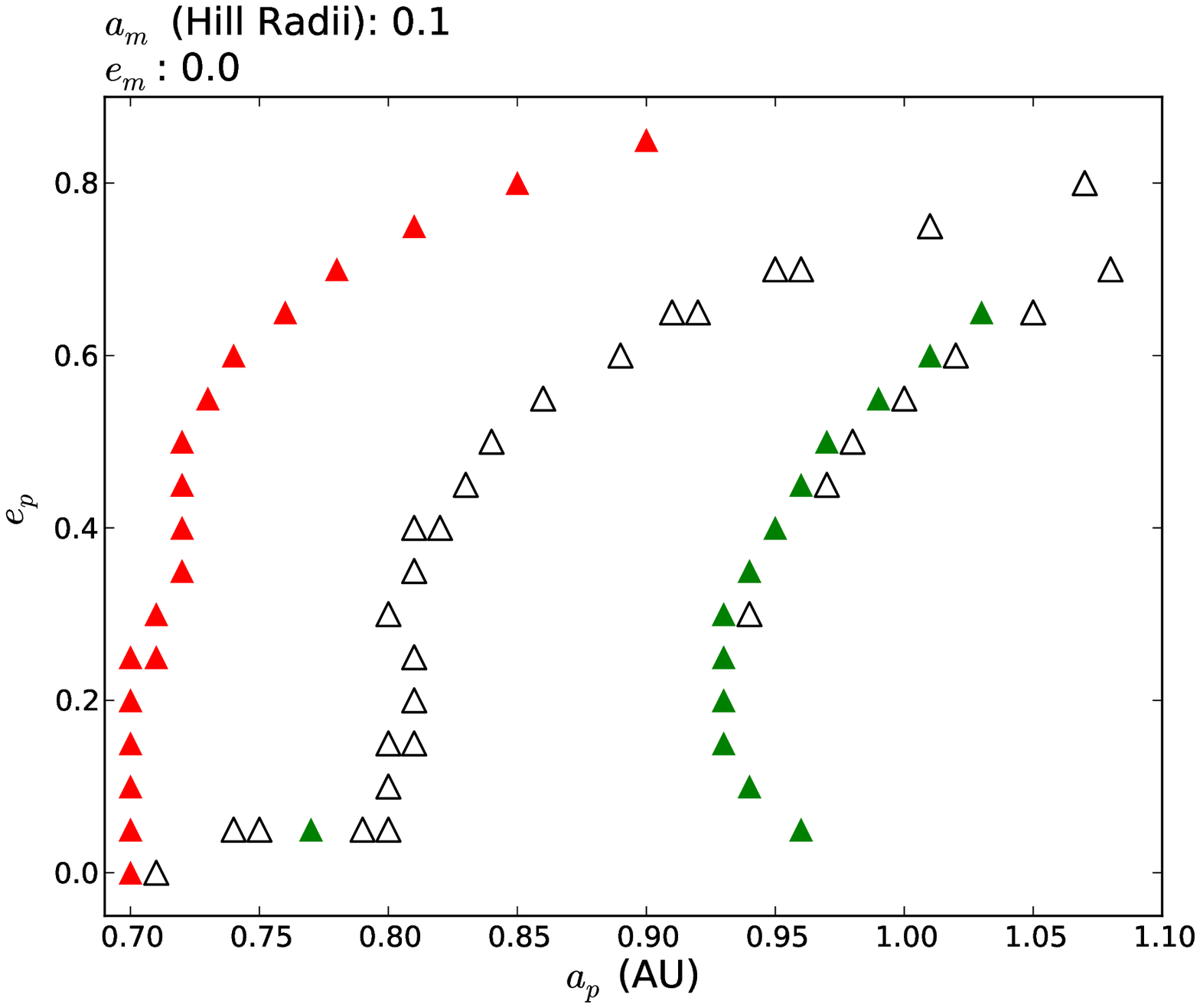} \\
\includegraphics[scale = 0.4]{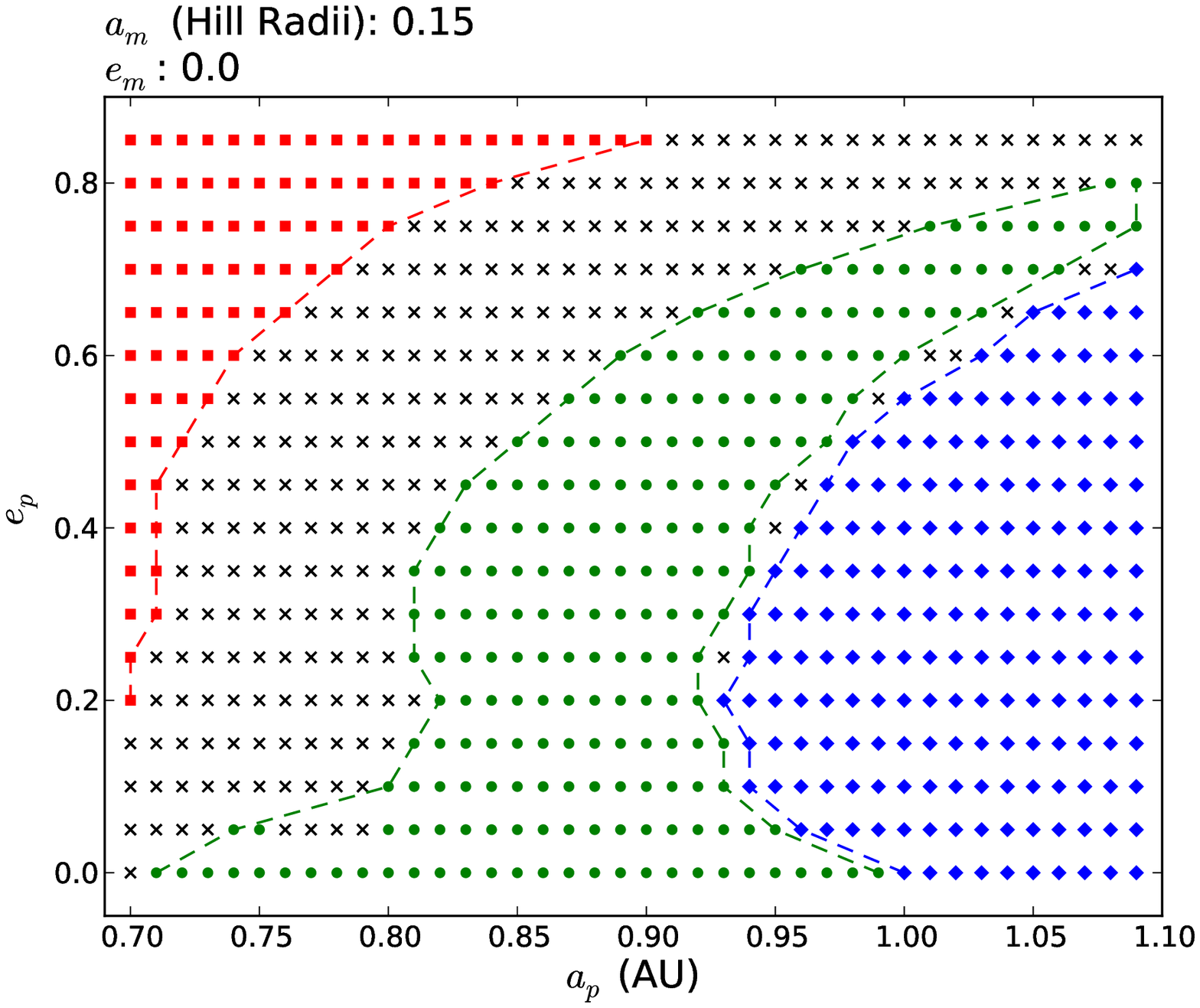} &
\includegraphics[scale = 0.4]{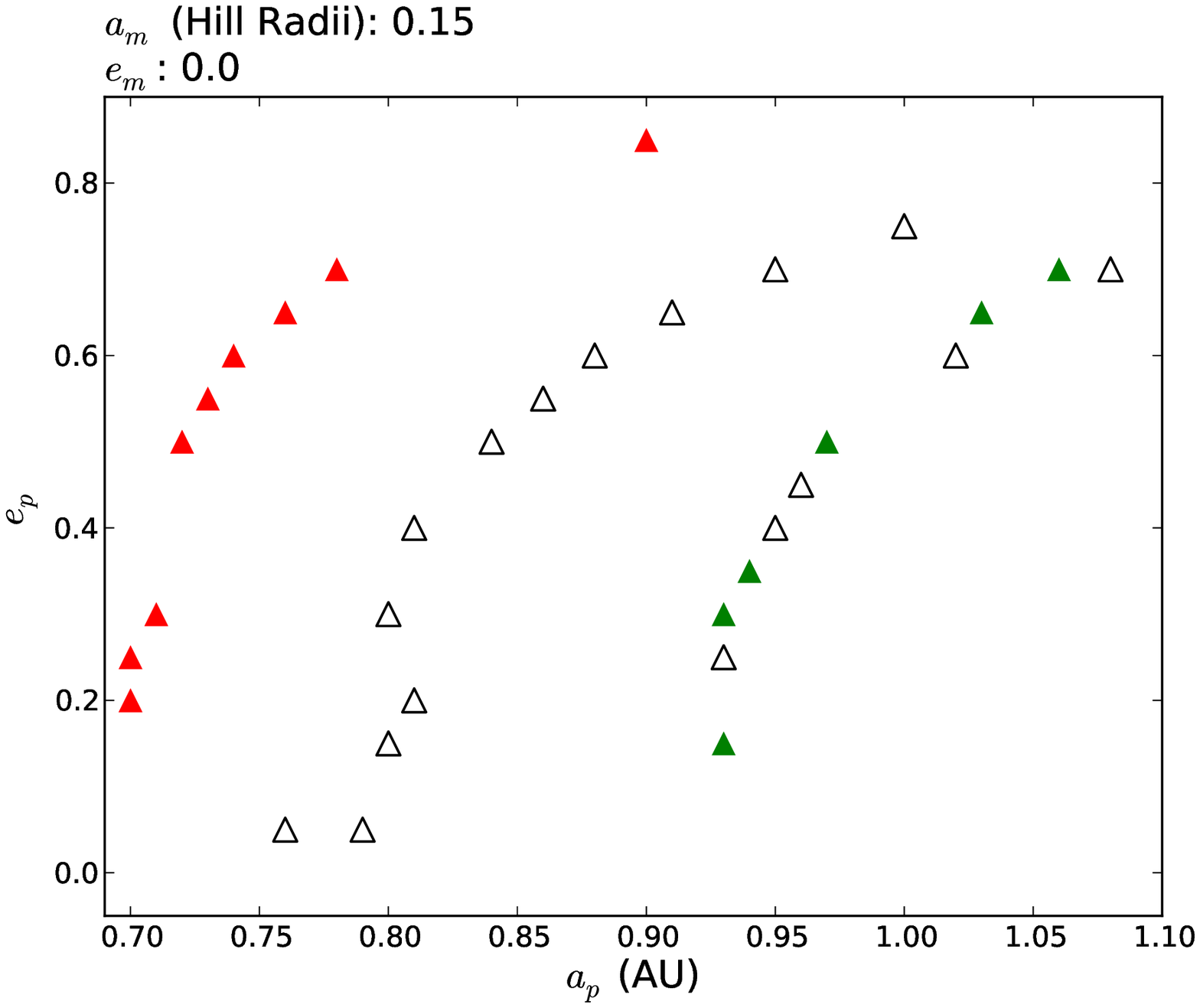} \\
\includegraphics[scale = 0.4]{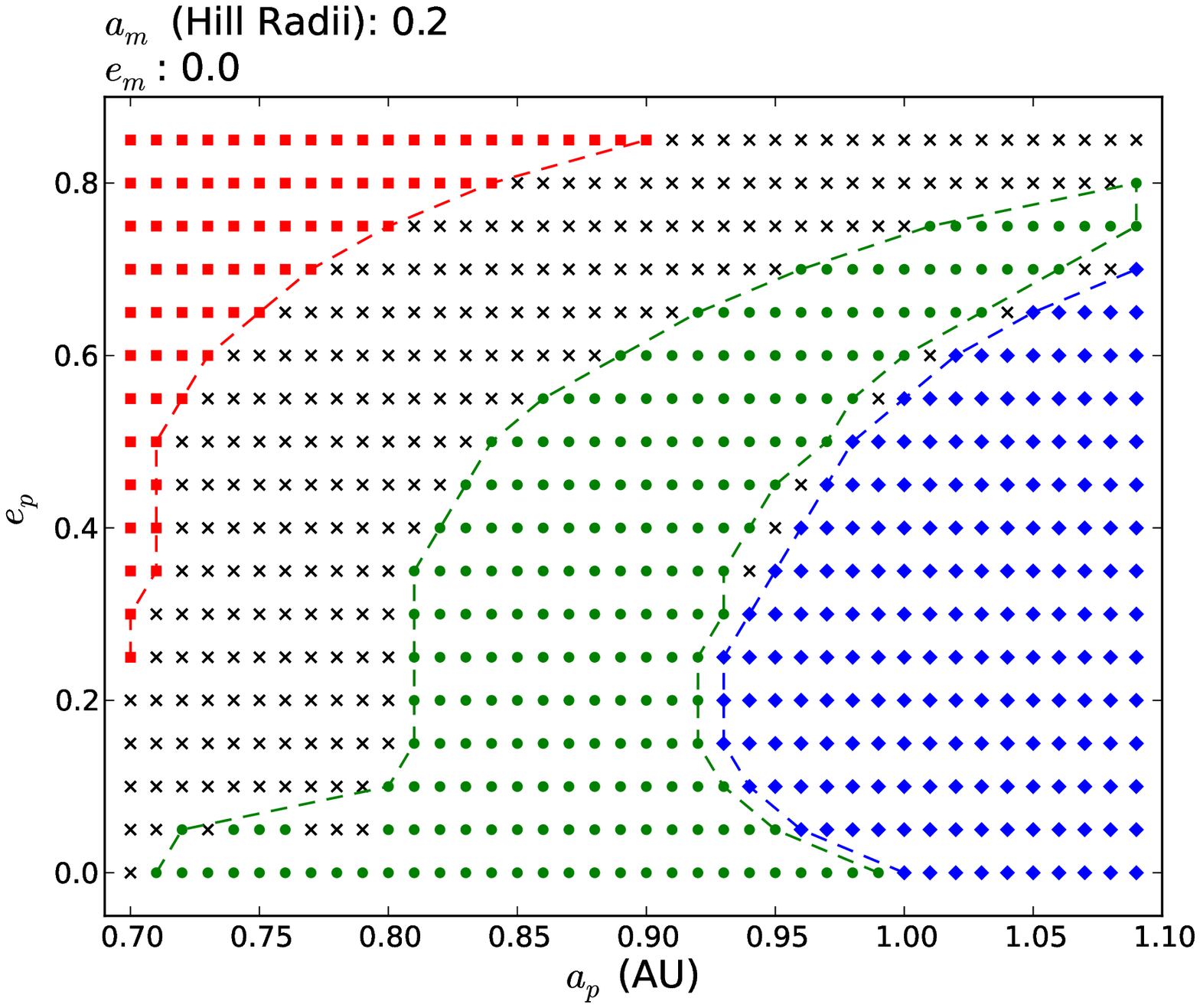} &
\includegraphics[scale = 0.4]{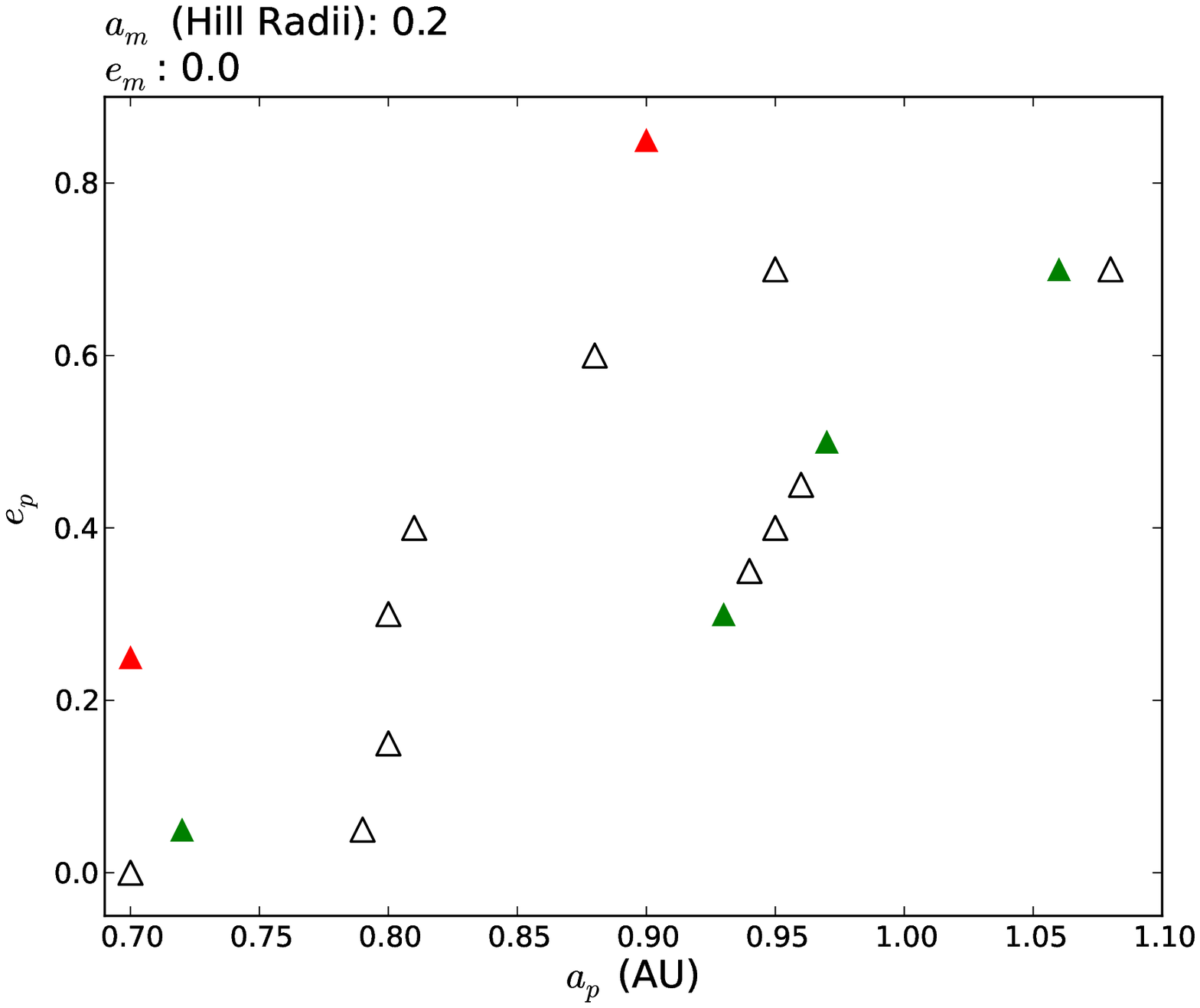} \\
\end{array}$
\caption{Left: The habitable zone, as a function of planet semi major axis $a_p$ and eccentricity $e_p$.  The moon's eccentricity is fixed at $e_m=0$, and the moon's semimajor axis $a_m$ is labelled above each plot, in units of local Hill Radius.  Each point represents a LEBM simulation; green circles designate simulations which return a habitable moon, red squares designate a hot moon, blue diamonds a cold moon, and crosses designate transient moons (where the definitions of each are given in section \ref{sec:habindex}).  Right: the relative change in habitability class as a result of adding planetary illumination.  Simulations where the habitability classification did not change as a result of illumination are not plotted. Upward triangles indicate that the new habitability classification is warmer, and colours indicate the classification received with planetary illumination. \label{fig:e0}}
\end{center}
\end{figure*}

\noindent By setting $e_m=0$, tidal heating is reduced to zero, and we can investigate the effects of planetary illumination without this extra complication.  Figure \ref{fig:e0} shows habitable zones calculated as a function of $a_p$ and $e_p$, where $a_m$ is fixed at a fraction of the local Hill Radius.  The left panels of Figure \ref{fig:e0} show the habitable zone calculated with planetary illumination, and the right panels compare simulations with and without illumination, plotting simulations where adding planetary illumination alters the habitability classification of the moon.

We can see that the morphology of the habitable zone for an exomoon on a circular orbit has important differences to that of a planet for the same $a_p$, $e_p$.  The outer boundary of the HZ (where worlds transition from warm to cold) moves inward at moderate values of $e_p$.  The inner boundary (where worlds transition from warm to transient) is also pushed inward at similar values of $e_p$ - in both cases, this is caused by the cooling effect of eclipses.

The overall shape of the HZ does not change as $a_m$ is changed, but it is clear that as expected, planetary illumination is most effective at low $a_m$.  The top right panel of Figure \ref{fig:e0} shows that the boundaries between each habitability classification is moved outwards in $a_p$ by 0.01 AU at all values of $e_p$ by adding planetary illumination.  As $a_m$ is increased, this effect diminishes, but the overall temperature of all moons increases slightly due to the extra source of energy.  This diminishing happens most rapidly at high values of $a_p$, where planetary illumination is least dominant.

We conducted runs with $dT=100$ K, and $dT=0$ K, and found no significant difference.   There were no changes in habitability classification as a result of changing $dT$, and the moons' surface temperatures changed very little.

\subsubsection{Adding Tidal Heating}

\begin{figure*}
\begin{center}$
\begin{array}{cc}
\includegraphics[scale = 0.4]{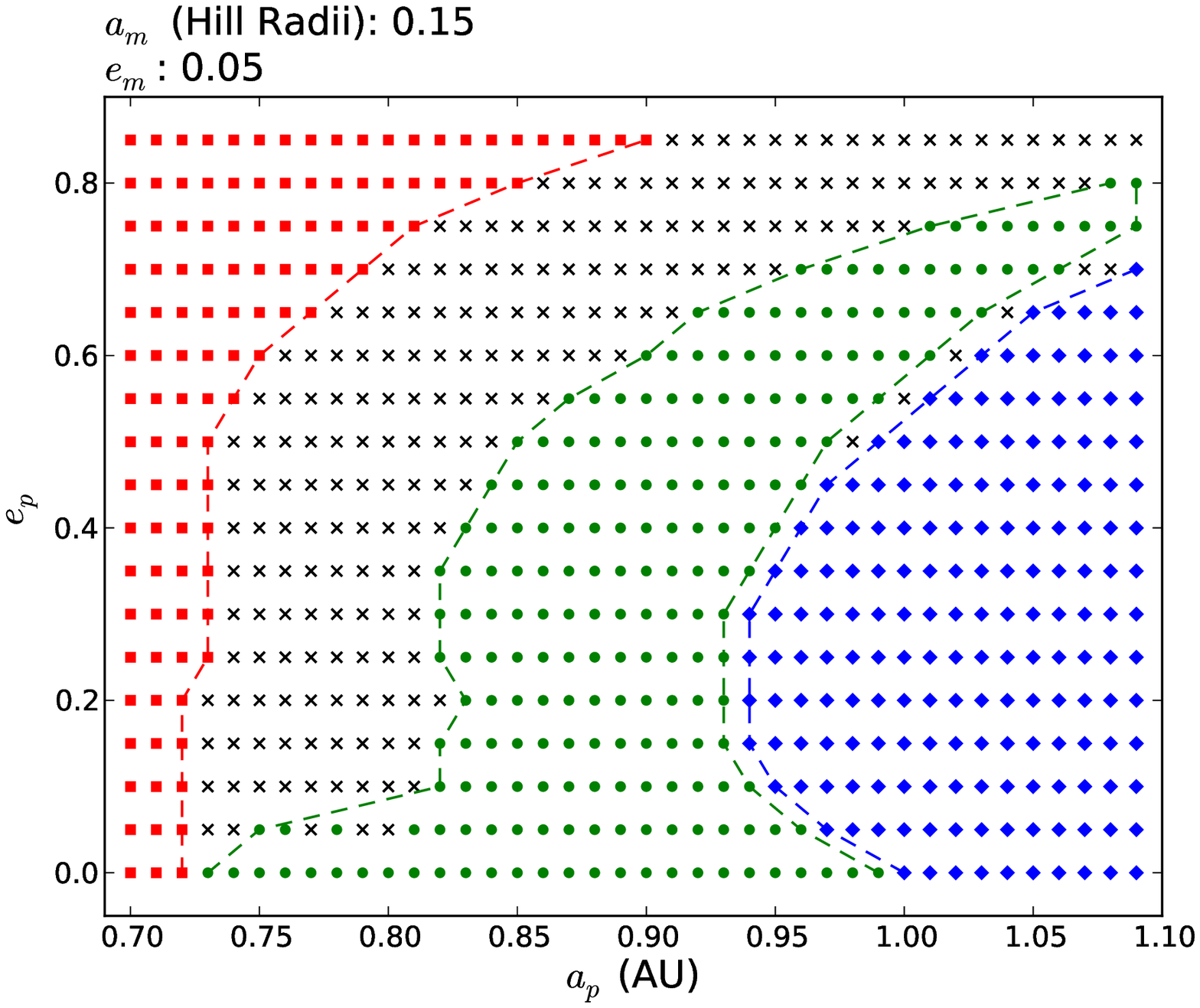} &
\includegraphics[scale = 0.4]{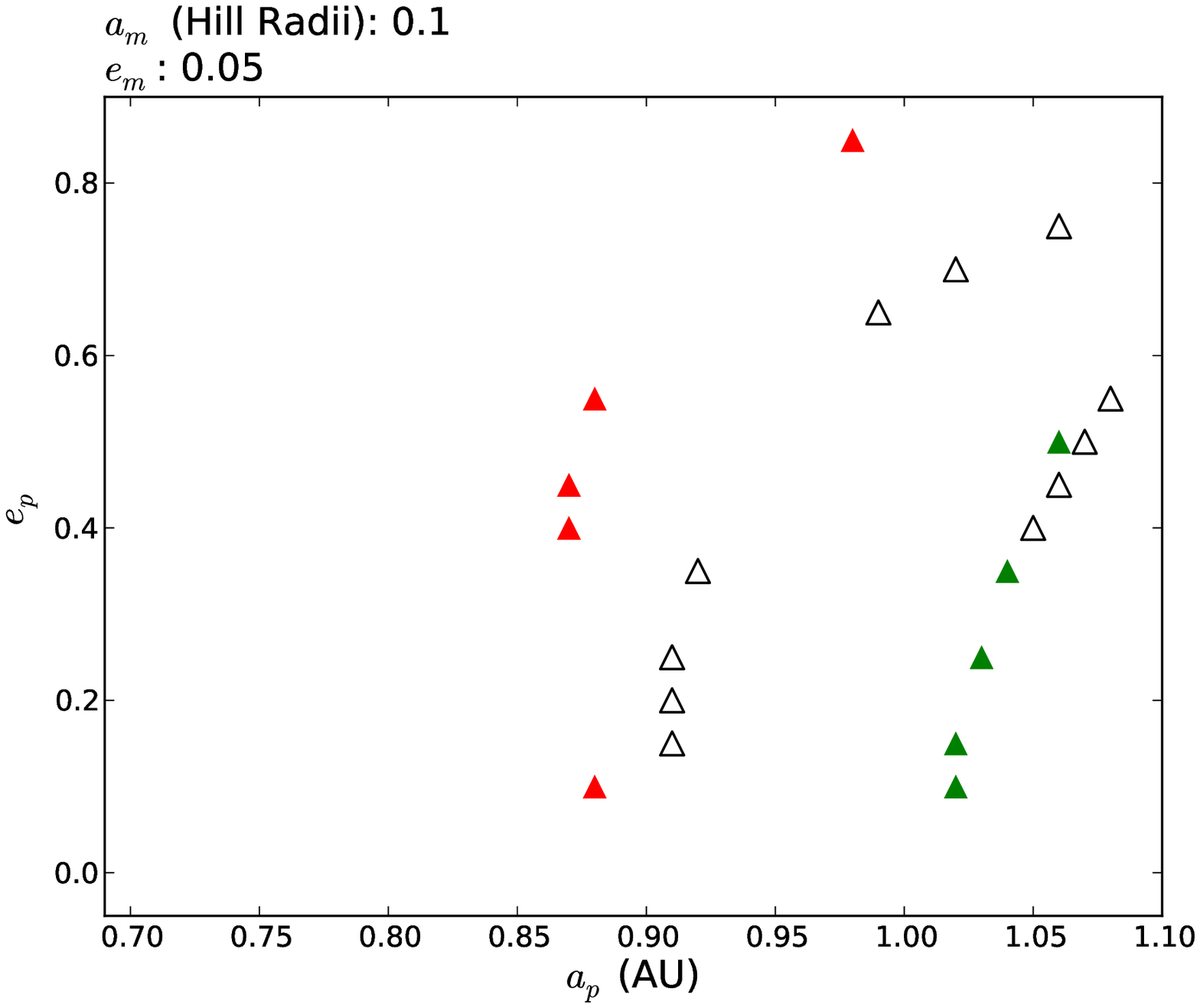} \\
\includegraphics[scale = 0.4]{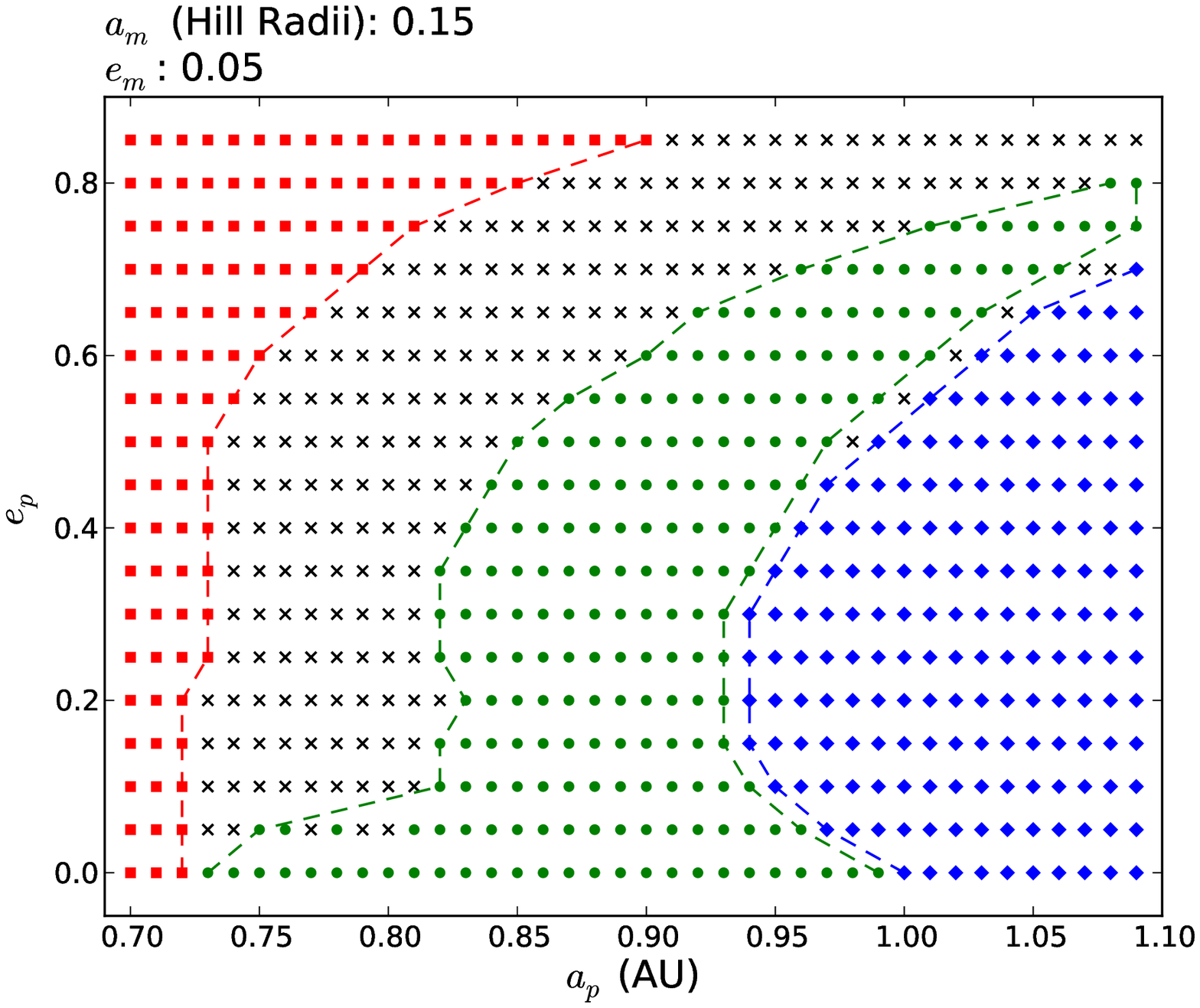} &
\includegraphics[scale = 0.4]{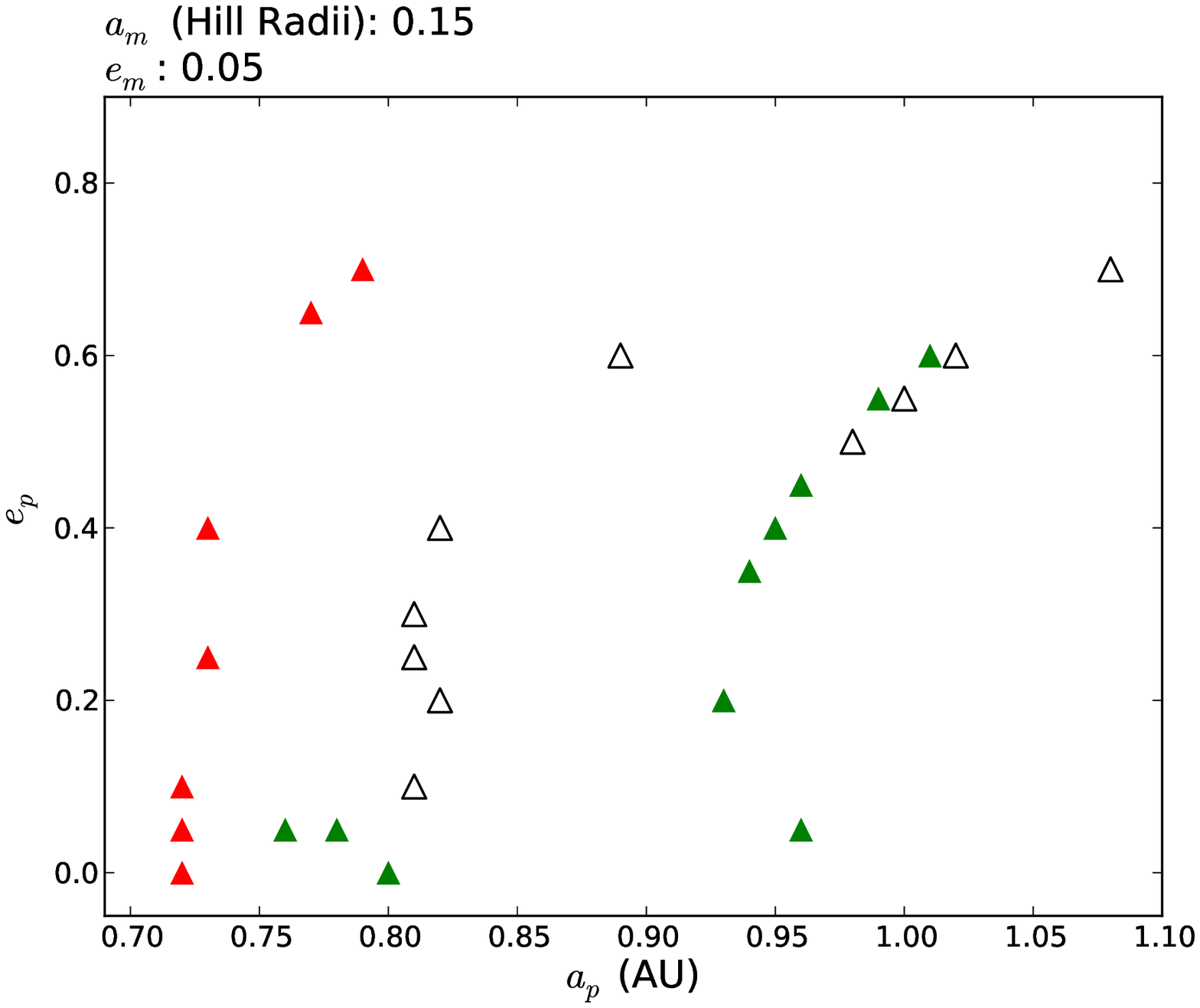} \\
\includegraphics[scale = 0.4]{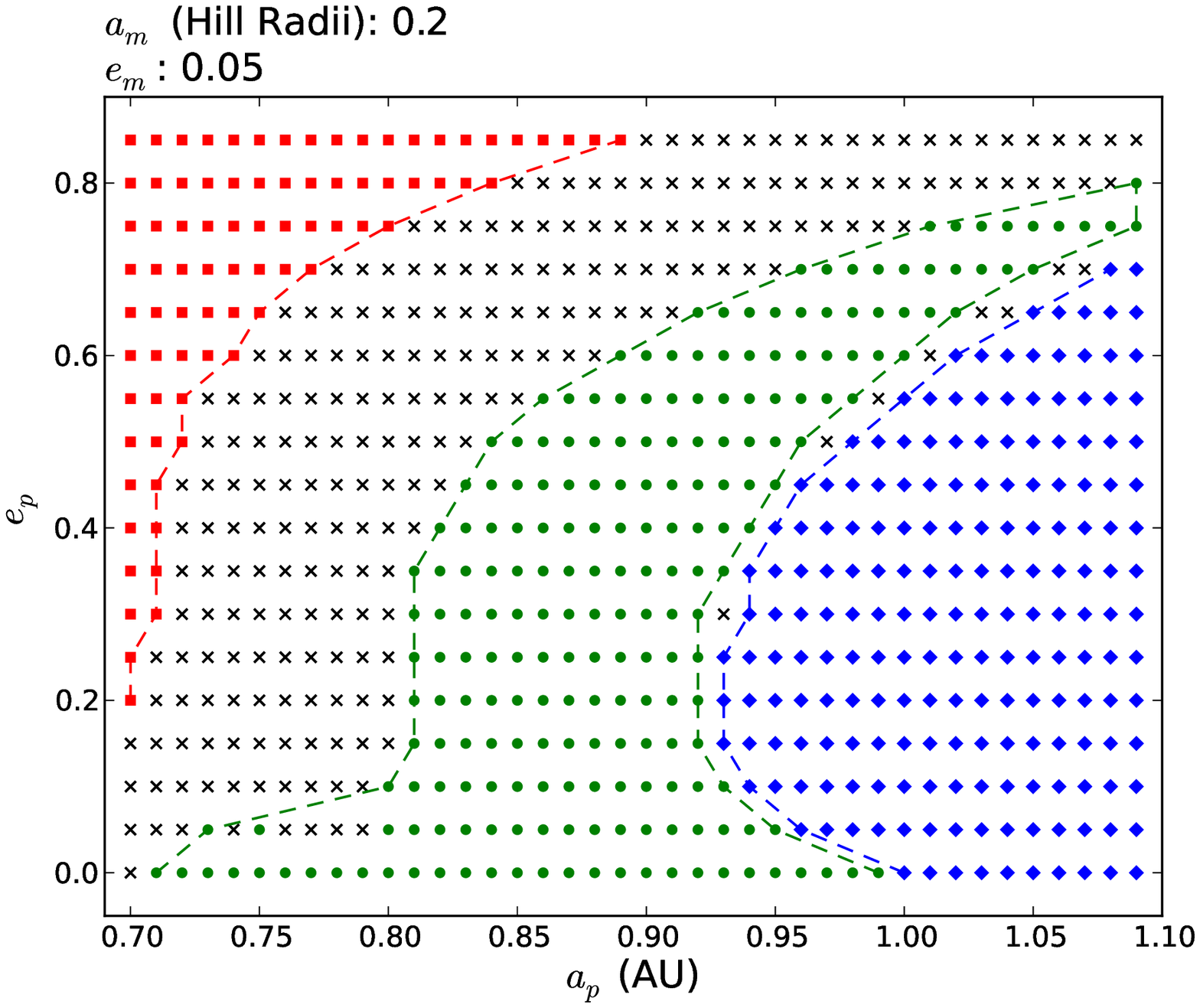} &
\includegraphics[scale = 0.4]{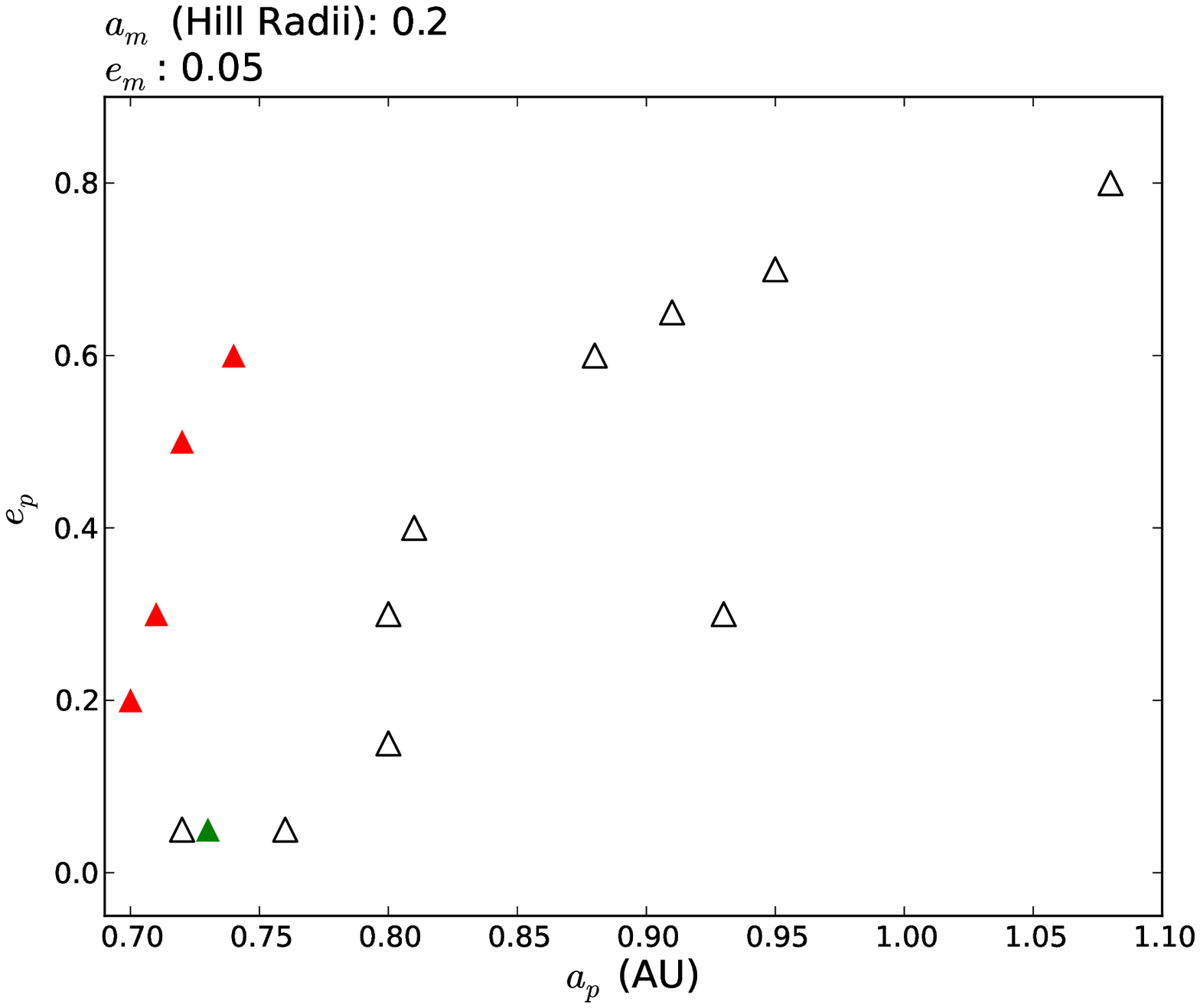} \\
\end{array}$
\caption{As Figure \ref{fig:e0}, but with the moon eccentricity now fixed at $e_m=0.05$. \label{fig:e0.05}}
\end{center}
\end{figure*}

\noindent By setting $e_m=0.05$, we now introduce a moderate amount of tidal heating.  Figure \ref{fig:e0.05} shows the same parameter space of simulations as the previous section.

At low $a_m$, tidal heating dominates the local energy budget, pushing the inner HZ boundary outwards by 0.1 AU in $a_p$.  The outer HZ boundary remains at a similar location, but the curvature of the boundary is reduced, making it closer to a vertical boundary at low values of $e_p$.  As $a_m$ is increased, the tidal heating is reduced until it is no longer a dominant energy source, and the HZ measured at $a_m=0.2$ Hill Radii is difficult to distinguish from that measured at $e_m=0$.  

With tidal heating now in the mix, planetary illumination exerts less of an influence on the HZ (as is evident in the right hand panels of Figure \ref{fig:e0.05}).  While adding illumination does push the HZ outward in $a_p$, the inner HZ boundary remains strongly governed by tidal heating.  Conversely, the outer boundary is more amenable to modification by adding planetary illumination.

Again, we find that the value of $dT$ has little appreciable effect on the EHZ.   However, eccentric moon orbits can also affect the eclipse rate, which depends on the moon's longitude of periapsis.  If a moon happens to be at apapsis during an eclipse, that eclipse can last somewhat longer and push the moon into a snowball state.  If the moon's host planet is at large $a_p$, it will be difficult to exit a snowball phase as the moon's albedo is large and the stellar flux is relatively low.  Planetary illumination alone cannot move a system away from snowball, and if $a_m$ is large, then tidal heating will also be ineffective.  This results in the bottom left panel of Figure \ref{fig:e0.05} having slightly more snowball results at moderate eccentricity than its counterpart in Figure \ref{fig:e0}.

\subsection{Exomoon Habitability as a Function of Moon Orbit}

\citet{Heller2013} define the ``circumplanetary habitable edge'' as a constraint on the minimum separation between the planet and moon.  Inside this separation, the effects of either planetary illumination, tidal heating (or both) precipitate a runaway greenhouse effect, rendering the moon uninhabitable.  It is typically referred to as an edge, as analytical calculations find no corresponding outer edge - the only limit on the habitability of Earthlike moons appears to be the orbital stability limit.  We have defined this stability limit as $a_m < 0.3 R_H$, which as we have mentioned is relatively conservative compared to other estimates, which demand that the orbital period of the moon be one ninth that of the planet (cf \citealt{Heller2012}).  Our constraint is equivalent to demanding that the orbital period of the moon be less than one tenth that of the planet.

\begin{figure*}
\begin{center}$
\begin{array}{cc}
\includegraphics[scale = 0.4]{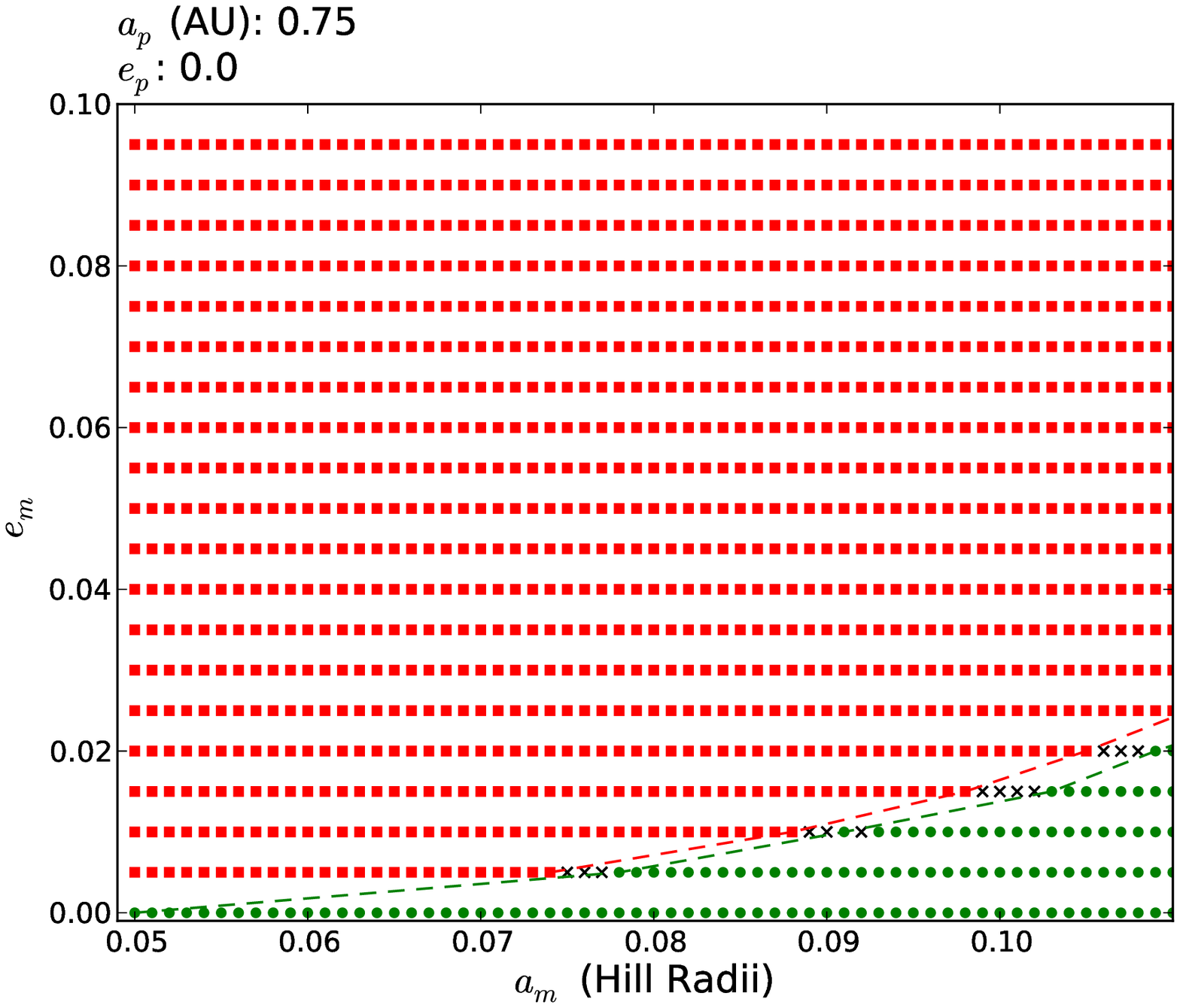} &
\includegraphics[scale = 0.4]{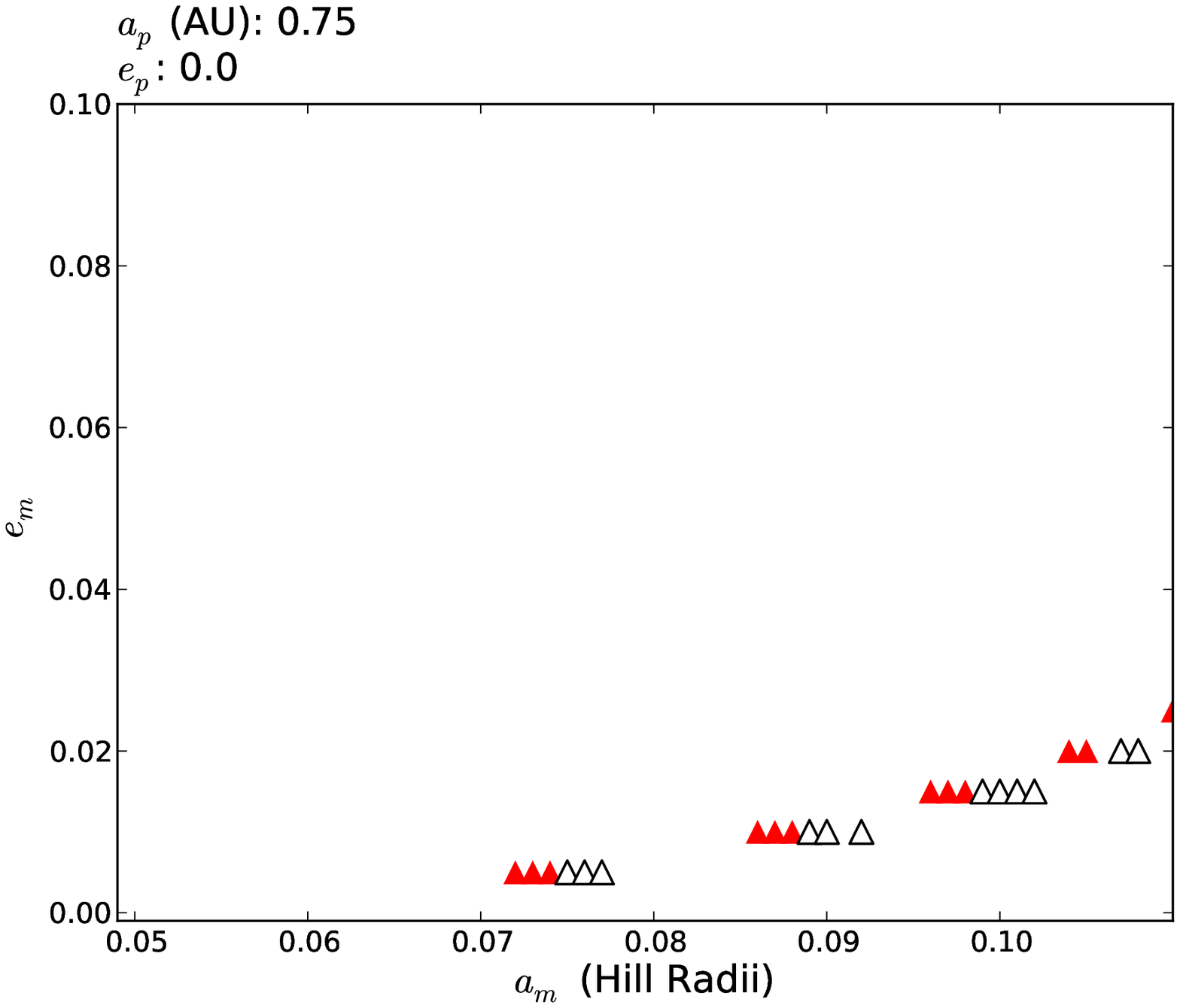} \\
\includegraphics[scale = 0.4]{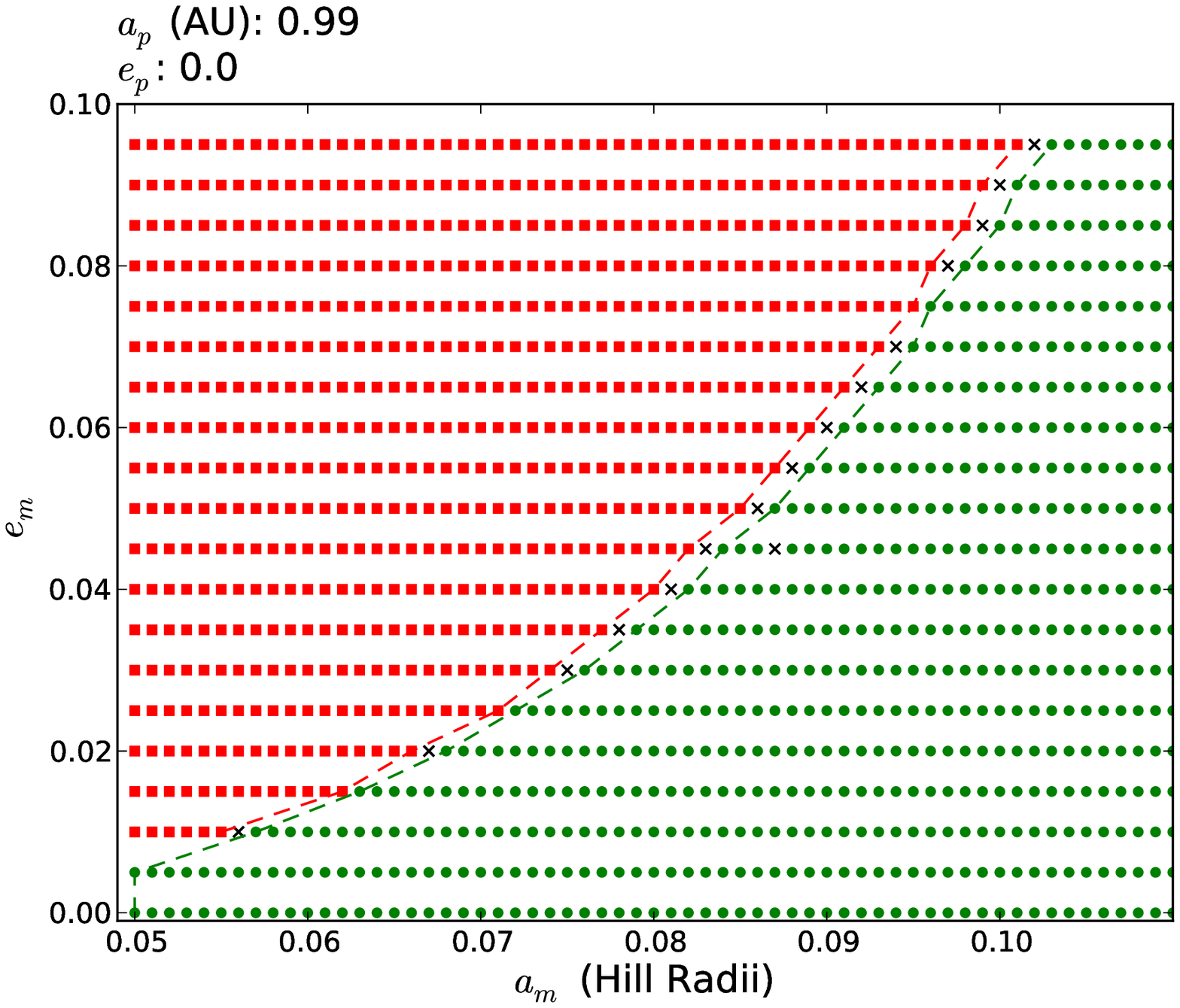} &
\includegraphics[scale = 0.4]{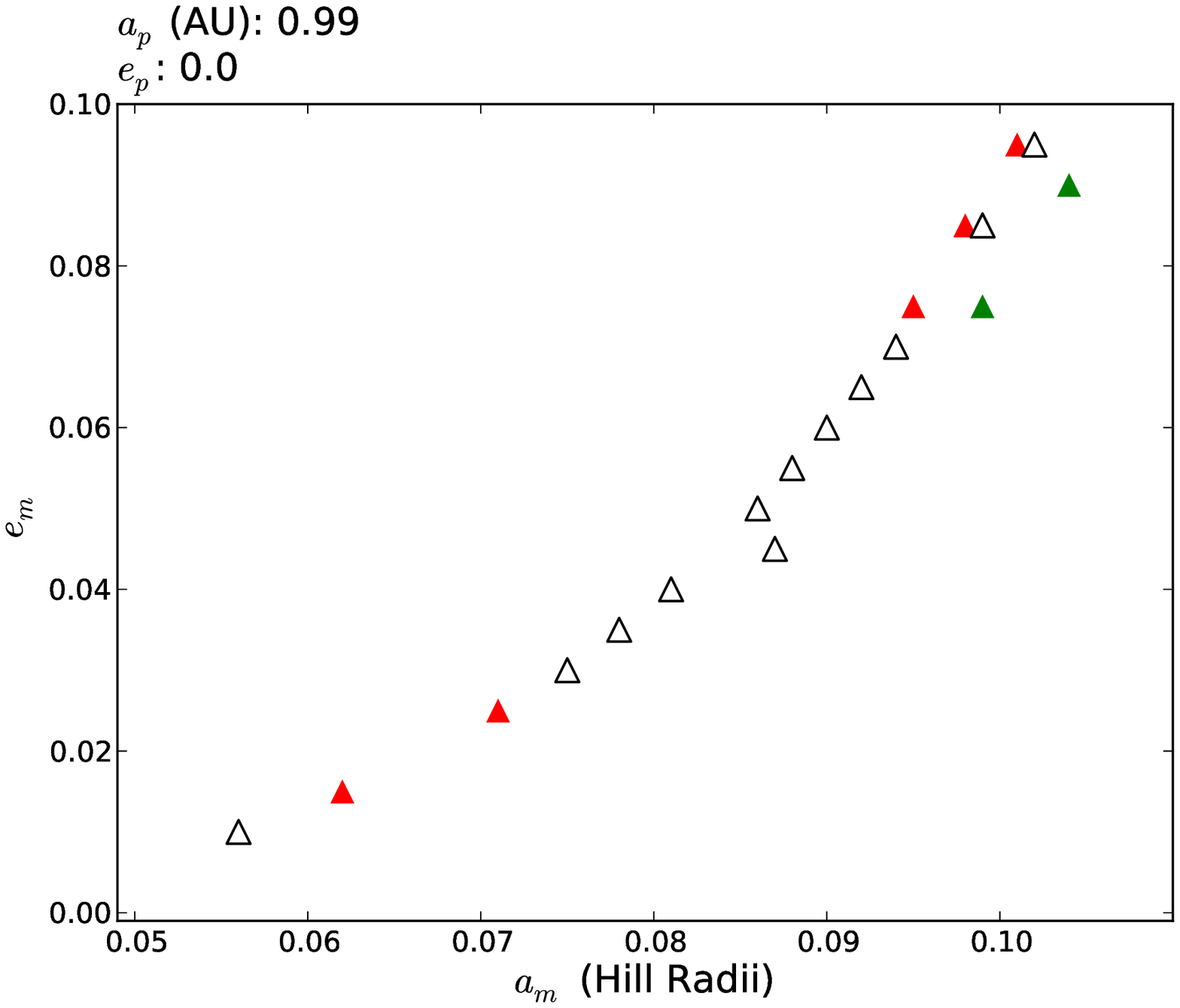} \\
\includegraphics[scale = 0.4]{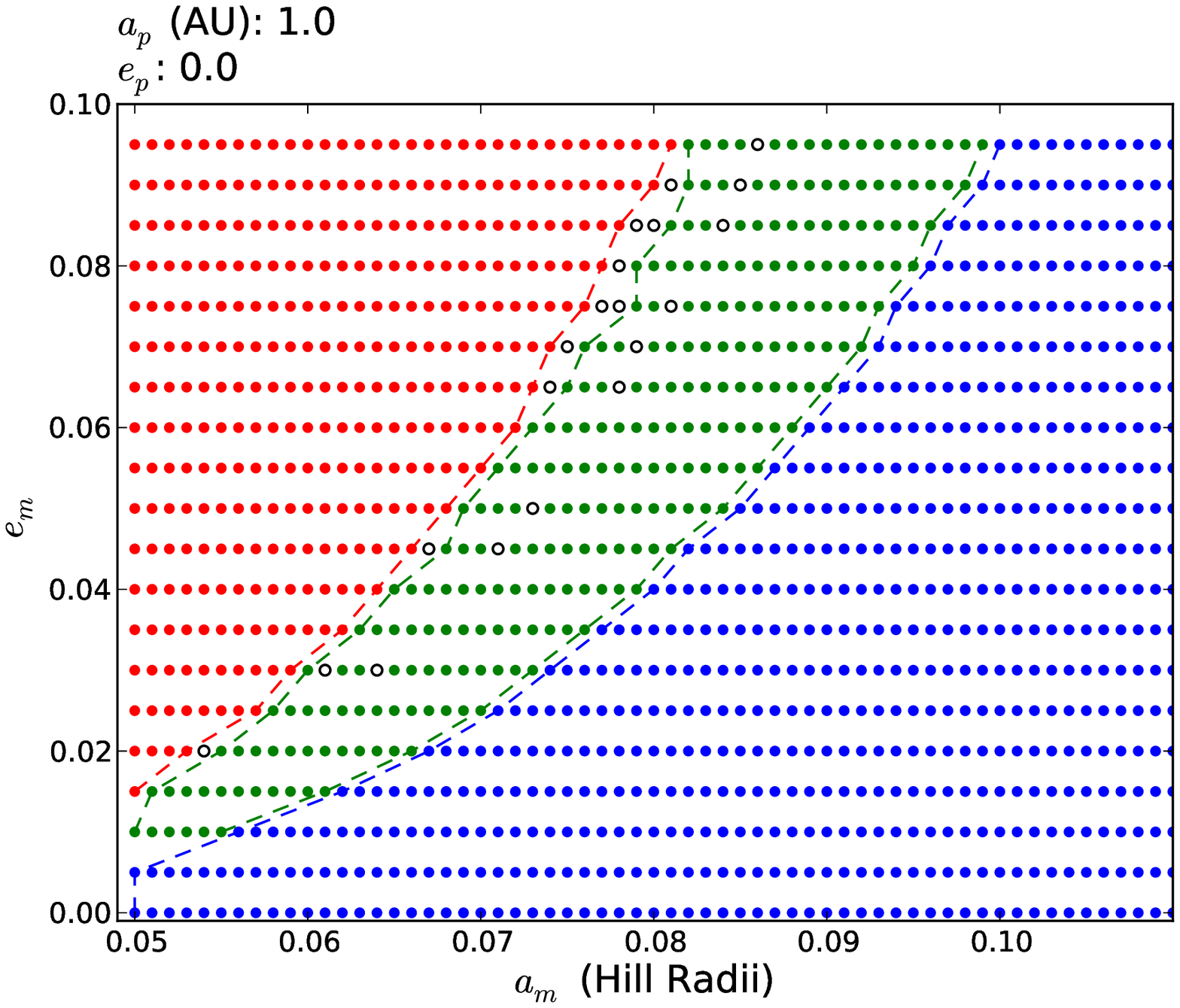} &
\includegraphics[scale = 0.4]{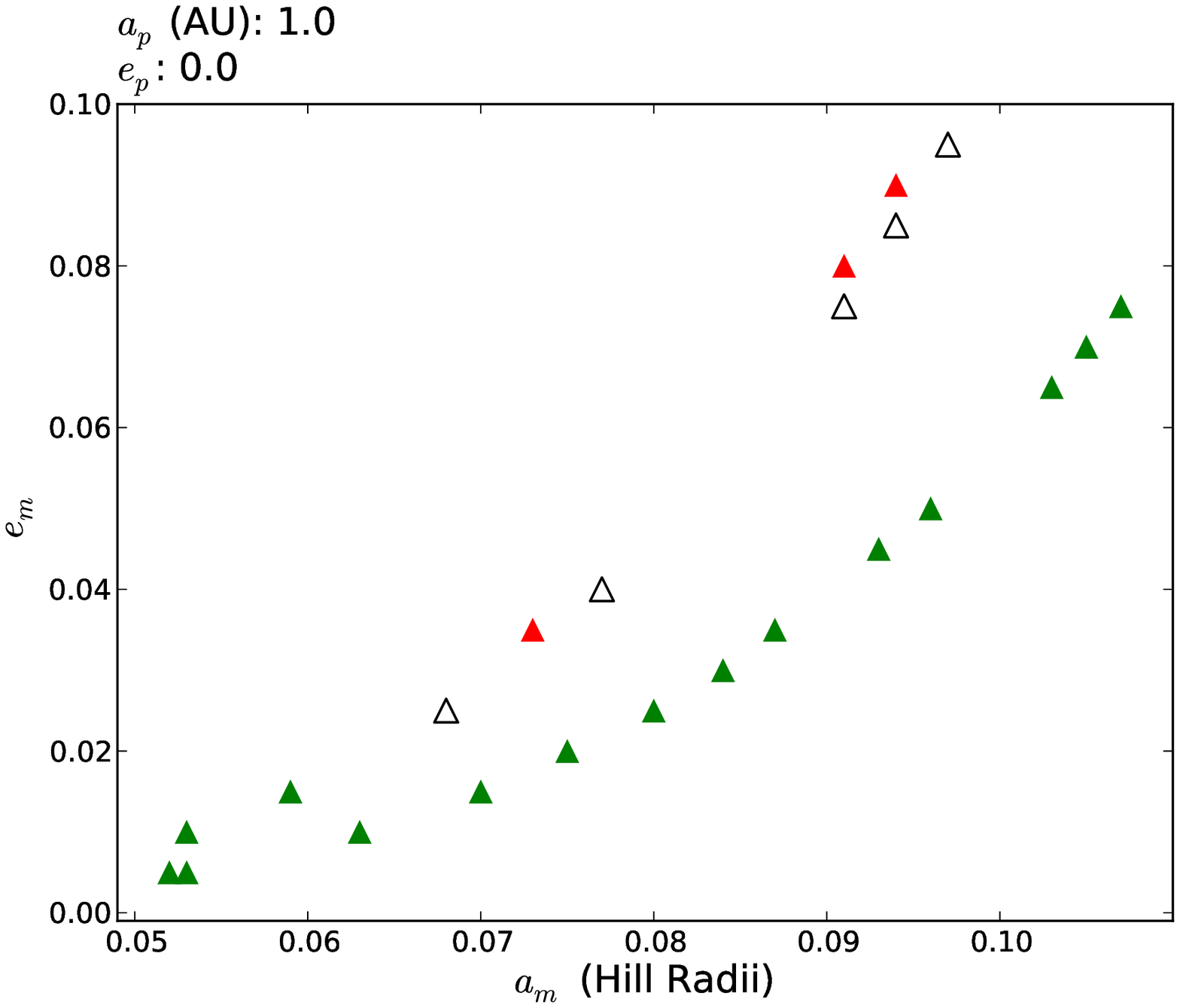} \\
\end{array}$
\caption{The circumplanetary habitable edge, as measured at a planetary semimajor axis of $a_p=0.75$ AU (top row), $a_p=0.99 $ AU (middle row) and $a_p=1AU$ (bottom row). The planetary eccentricity is fixed at $e_p=0$.  As with previous figures, the left column shows the sets of LEBM simulations run with planetary illumination active, and the right column compares simulations run with and without planetary illumination.  The colours/shapes of the points corresponds to the same classification as previous figures. \label{fig:habedge}}
\end{center}
\end{figure*}

With planetary illumination now part of the LEBM, we can now investigate this habitable edge using climate modelling for the first time.  We fix $a_p$, and set $e_p=0$, and vary $a_m$ and $e_m$.  Figure \ref{fig:habedge} shows how the circumplanetary habitable edge changes with $a_p$.  

The top row ($a_p=0.75$ AU) shows that deep inside the habitable zone defined by $a_p$ and $e_p$, there is only one habitable edge.  While the simulations shown here extend only to $a_m=0.12$ Hill Radii, simulations were run out to $a_m=0.3$ Hill Radii and remained habitable for all values of $e_m$.  Adding illumination pushes this edge outwards by around 0.006 Hill Radii at most values of $e_m$ (top right panel of Figure \ref{fig:habedge}). 

The middle row ($a_p=0.99$ AU) is near the edge of the habitable zone defined by $a_p$ and $e_p$, but we still see evidence of only one habitable edge.   Moving the planet outwards has reduced its equilibrium temperature, and hence its thermal flux.  The reflected stellar flux has also decreased.  The inner habitable edge has moved inwards (as a function of Hill Radius).   Equally, the Hill Radius increases linearly with $a_p$, so the absolute distance of the moon from the planet will also be larger.  Increasing $a_p$ from 0.75 to 0.99 will increase the Hill Radius by a factor of 1.3.  The inner edge has moved inwards by a factor of around 1.6, so there is an effect attributable to physical processes, rather than simply a choice of units.  At our minimum $a_m$ of 0.05 $R_H$, moons with non-zero $e_m$ can now be habitable, showing how much tidal heating has been reduced.  As the flux from planetary illumination has been reduced by increasing $a_p$, its effect on the habitable edge is minimal, as can be seen in the middle right panel of Figure \ref{fig:habedge}.  

The bottom row ($a_p=1$ AU) is on the edge of the habitable zone defined by $a_p$ and $e_p$, and we now see a circumplanetary habitable zone, rather than a single habitable edge.  Large values of $a_m$ now result in snowball moons, due to relatively long eclipses.   This outer habitable edge resides well within the orbital stability limit, and in contrast to the $a_p=0.99$ AU suite, is more sensitive to planetary illumination.  The total energy budget (stellar flux, planetary flux and tidal heating) is extremely low, such that relatively small amounts of planetary flux results in a significant effect.  This is evident in the lack of clear boundaries between simulations classified as warm and transient in the bottom left panel of Figure \ref{fig:habedge}.  The bottom right panel of Figure \ref{fig:habedge} shows that adding illumination pushes out the outer habitable edge by 0.001 Hill Radii.  We confirm that due to the high dimensionality of the EHZ, exomoon orbits that appear to spend some or all of their time outside the planetary HZ can produce habitable exomoons \citep{Hinkel2013, Heller2014}.

\section{Discussion}\label{sec:discussion}

% Better Tidal Heating
% Wavelength Dependence of Albedo
% Classification of moons...
%  Mass dependence of 

Use of a 1D LEBM necessitates a number of assumptions.  Some (like assuming the moon's rotation period is relatively rapid) are necessary in order to reduce the problem successfully to one dimension.  Others are necessary for simplicity's sake, so that the effect of adding physical processes to the system can be completely understood.

One important example is the absence of a carbonate-silicate cycle.  Allowing the partial pressure of $CO_2$ to respond to geodynamical processes, such as volcanic outgassing and silicate weathering, produces a feedback system that can prevent the snowball effect from taking hold when insolation is low \citep{Williams1997a}.  LEBM simulations of planets that use the carbonate-silicate cycle typically allow the $CO_2$ pressure to vary on timescales of an orbital period, but it is clear that the cycle will not operate in precisely the same fashion for Earthlike exomoons.  From studies of the cycle on tidally-locked planets \citep{Joshi1997,Edson2012}, we can speculate that the cycle's behaviour will vary depending on the moon's rotation and orbital periods around the planet, as well as the distribution of oceans and continents, but further work is required to establish this.

The distribution of continents also has implications for the tidal heating of the moon's surface.  We have assumed that tidal heating is uniformly distributed, but this may not necessarily be true.   In the case of Io, it seems that tidal heating has maxima at the poles, and a minimum at the subplanetary point \citep{Segatz1988}.  Naturally, studies of dry, highly volcanic moons like Io have limited application to Earthlike moons, especially as tidal forces will affect land and ocean differently according to their elastic rigidities.  We have fixed the ocean fraction of the moon surface at 0.7, but our model does not incorporate this into calculating the energy flux from tidal heating.  Future work should consider more carefully how to implement tidal heating, for example using a continuous phase lag (CPL) model \citep{Ferraz-Mello2008}.  This would allow a more rigorous calculation of heating due to effects including orbital circularisation and obliquity erosion.  Such models may also allow a more rigorous latitudinal distribution of heat, incorporating the potential for tidal hotspots (see \citealt{Heller2013b} and references within).

We have noted that the EHZ is a multidimensional manifold, but the observant reader will note that its dimension is likely to be even higher than stated here.  If planetary illumination and tidal heating help shape the circumplanetary habitable zone, then the planet's structural parameters - its mass and radius - will also shape the EHZ.  For clarity, we have only studied a $1 \mjup$, $1 \rjup$ planet in this work, but future work should explore how these parameters affect the EHZ.  Also, we have assumed the star-planet-moon system is coplanar, with the moon's orbit being prograde.  Perfectly aligned systems like these are unlikely, and our previous work has shown that inclination has an important effect on exomoon climates \citep{Forgan_moon1}.  It is clear that changing inclination will change the frequency of eclipses, which will have consequences for the circumplanetary habitable zone.

This work considers Earthlike moons only, by necessity.  The value of the diffusion constant $D$ is calibrated from fiducial models of the Earth-Sun climate to reproduce the correct latitudinal temperature distribution (cf  \citealt{Spiegel_et_al_08,Vladilo2013}).  It would be useful to extend the limits of this model beyond Earthlike moons to represent potentially habitable moons we already know of - the Galilean moons and Titan.  Figure 4 of \citet{Heller2013b} is a good demonstration of how this can be done for analytical calculations using globally-averaged radiative flux.

In principle, LEBMs could be constructed to simulate the temperature evolution of these moons, but this would require new functions to describe the albedo, heat capacity and optical depth, as well as a correctly calibrated $D$.  In the case of Titan, a variety of climate models currently exist, from globally averaged models \citep{Zahnle2014} to full global circulation models (GCMs) \citep{Lora2013} - these models would be the first place to start when attempting to create a Titan LEBM. 

The presence of an outer circumplanetary habitable edge in these simulations is in contrast to previous work, but it does echo analytical calculations made by \citet{Heller2014}, which demonstrate that moons orbiting planets outside the traditional planetary HZ can maintain habitable surface temperatures through tidal heating and illumination.  The nature of this outer edge will depend on the moon's ability to forestall the snowball mechanism.  The primary means by which the moon could prevent a snowball event is adjustment of $CO_2$ levels using the carbonate-silicate cycle, another reason why future models must include this physical process.

The propensity for moons to be eclipsed by their host planet depends on the inclination of the moon relative to the plane defined by the star-planet system.  In these simulations, we have focused on the perfectly coplanar case, where the moon's inclination relative to the planet's orbit $i_m=0^\circ$.  A non-zero $i_m$ will in general decrease the eclipse rate, as will increasing $a_m$ (see e.g. \citealt{Heller2012}), although increasing $a_m$ will also lengthen the duration of the eclipse.  Again, the planet's structural properties, in particular its physical radius, will affect the eclipse timescale, and hence the shape of the outer habitable edge, once more highlighting the need for future work to probe exomoon habitability as a function of planet mass and radius.

It is also clear that circumplanetary habitability will depend on how albedo is modelled.  The stellar and planetary flux will generally arrive at the moon's surface in different wavelength bands, and as a result we should expect the atmosphere's response to also be a function of wavelength.  Our assumption of a single albedo is therefore over-simplified.  A more appropriate model may be the two albedo approach used by \citet{Heller2013b}.

Finally, we have neglected the role of gravitational perturbations in the climate of these moons.   Fixed Keplerian orbits are convenient, but do not reflect (for example) the tidal evolution of the star-planet-moon system (e.g. \citealt{Laskar1993, Sasaki2012}).  Gravitational perturbations of other bodies upon the Earth's motion result in Milankovitch cycles \citep{Berger2005,Spiegel2010}.  Coupling the LEBM with algorithms to calculate the evolution of the moon's spin and orbit would allow an investigation of how a moon's orbit and climate are coupled.

\section{Conclusions }\label{sec:conclusions}

\noindent We have continued our work in 1D latitudinal energy balance models (LEBMs) of exomoon climates, adding in the extra radiation source that is the host planet.  The model now incorporates stellar insolation, planetary insolation from thermal blackbody radiation and reflected starlight, tidal heating, eclipses of the star by the planet, diffusion of heat across latitudes and atmospheric cooling.  It is the first climate model of its kind to contain all the primary sources and sinks of energy that dictate the radiative energy budget of exomoon systems.

We have used this upgraded LEBM to explore the exomoon habitable zone (EHZ), in four dimensions: the planet semimajor axis $a_p$ and eccentricity $e_p$, and the moon semimajor axis $a_m$ and eccentricity $e_m$.  In terms of $a_p$ and $e_p$, the EHZ overlaps most of the planetary habitable zone, but is extended marginally outward in $a_p$ by the effect of planetary illumination.  

In terms of $a_m$ and $e_m$, we find an inner circumplanetary ``habitable edge'' produced by tidal heating and planetary illumination, in line with previous studies \citep{Heller2013,Heller2013b}.  However, we also find evidence for an outer circumplanetary habitable edge, defined by eclipses.  This is found for large values of $a_p$, and exists at sufficiently low $a_m$ to be well within the orbital stability limit typically used as an outer habitability boundary.

To summarise, these models suggest that exomoon climates exhibit even more complex behaviour than was originally believed, and more work is required to determine under what circumstances outer circumplanetary habitable edges exist.

\section*{Acknowledgments}

\noindent DF gratefully acknowledges support from STFC grant ST/J001422/1.  The authors would like to thank the referee, Ren\'e Heller, for insights and comments which greatly improved this manuscript.

\bibliographystyle{mn2e} % (must include a bibliography style)
\bibliography{exomoon_irr}

\appendix

\label{lastpage}

\end{document}